\documentclass[a4paper,twocolumn,10pt,arxiv,accepted=2020-06-16]{quantumarticle}
\pdfoutput=1

\usepackage[utf8]{inputenc}
\usepackage[english]{babel}
\usepackage[T1]{fontenc}
\usepackage{hyperref}
\usepackage[rgb, dvipsnames]{xcolor}
\usepackage{bm}
\usepackage{amsmath}
\usepackage{amssymb}
\usepackage{amsfonts}
\usepackage{mathbbol}
\usepackage{graphicx}

\usepackage{setspace}

\usepackage{listings}
\definecolor{background}{rgb}{0.94,0.95,0.96}
\newcommand{\floorj}{\lfloor j\rfloor}
\global\long\def\ket#1{|#1\rangle}
\global\long\def\bra#1{\langle#1|}

\newcommand{\bq}{{\mathbf q}}
\newcommand{\bJ}{{\bf J}}
\newcommand{\hbq}{\hat{{\bf q}}}
\newcommand{\boldmu}{\boldsymbol{\mu}}
\newcommand{\bn}{\boldsymbol{\nu}}
\renewcommand{\leq}{\leqslant}
\renewcommand{\geq}{\geqslant}
\def\tr{\mathrm{tr}}
\def\idmat{\mathbb{1}}
\def\unitS{\mathcal{S}^{2}}

\makeatother

\begin{document}

\title{\parbox{\textwidth}{Optimal Detection of Rotations about Unknown Axes by Coherent and Anticoherent States}}

\author{John Martin}
\affiliation{Institut de Physique Nucléaire, Atomique et de Spectroscopie, CESAM, University of Liège, B-4000 Liège, Belgium}
\email{jmartin@uliege.be}
\orcid{0000-0003-0804-959X}
\author{Stefan Weigert}
\affiliation{Department of Mathematics, University of York, UK-York YO10 5DD, United
Kingdom}
\email{stefan.weigert@york.ac.uk}
\orcid{0000-0002-6647-3252}
\author{Olivier Giraud}
\affiliation{Université Paris-Saclay, CNRS, LPTMS, 91405 Orsay, France}
\email{olivier.giraud@universite-paris-saclay.fr}

\maketitle

\begin{abstract}
Coherent and anticoherent states of spin systems up to spin $j=2$
are known to be optimal in order to detect rotations by a known angle
but unknown rotation axis. These optimal quantum rotosensors are characterized
by minimal fidelity, given by the overlap of a state before and after
a rotation, averaged over all directions in space. We calculate a
closed-form expression for the average fidelity in terms of anticoherent
measures, valid for arbitrary values of the quantum number $j$. We
identify optimal rotosensors (i) for arbitrary rotation angles in
the case of spin quantum numbers up to $j=7/2$ and (ii) for small
rotation angles in the case of spin quantum numbers up to $j=5$.
The closed-form expression we derive allows us to explain the central
role of anticoherence measures in the problem of optimal detection of rotation angles for arbitrary values of $j$. 
\end{abstract}

\section{Introduction and main result}

Historically, advances in measurement techniques often are the reason
for physics to progress. Over time,\emph{ metrology} has developed
as a subject of its own, especially in the context of defining standard
units of measurement for physical quantities.

Quantum theory provides new perspectives on measurements, ranging
from fundamental limitations on measurements \cite{heisenberg27},
new opportunities \cite{gio11} as well as technical challenges
and even philosophical quagmires \cite{busch+2016}. From a practical
point of view, quantum information science requires ever better control
of microscopic systems and, hence, measurements which are as accurate
as possible. More specifically, quantum metrology \cite{Nawrocki19}
aims at finding bounds on the achievable measurement precision and
at identifying states which would be optimal for quantum measurements or other specific tasks. The optimal transmission of a Cartesian frame~\cite{Per01} or the efficient detection of inhomogeneous magnetic fields~\cite{Hak20} are typical examples. While the classical Cram\'er-Rao theorem \cite{Rao45, Cra46} provides
a lower bound on the variance of random estimators by means of the Fisher
information, its quantum-mechanical counterpart provides bounds for quantum
parameter estimation theory \cite{Hel76}. The quantum Cram\'er-Rao
bound is expressed as the inverse of the quantum Fisher information, which can be geometrically interpreted as the (Bures) distance between two quantum states differing by an infinitesimal amount in their parameter \cite{Hub92, Hub93}. It provides lower bounds on the variance of any quantum operator whose measurement aims at estimating the parameter. Optimal measurement is achieved by maximizing the quantum Fisher information over parameter-dependent states.

The quantum Cram\'er-Rao bound was calculated for instance in the reference
frame alignment problem \cite{Kol08}. This problem involves estimating
rotations about unknown axes. It has been shown in~\cite{Goldberg18} that spin states with vanishing spin expectation value and isotropic variances of the spin components are valuable for estimating such rotations, as they saturate the quantum Cram\'er-Rao  bound for \emph{any} axis. Also, recently, the problem of characterizing a rotation about an unknown direction encoded into a spin-$j$ state has been considered in~\cite{MoC19}. 

In this paper, we are interested to determine whether a quantum system
has undergone a rotation $R_{\mathbf{n}}(\eta)$ by a \emph{known}
angle $\eta$ about an \emph{unknown} axis $\mathbf{n}$. Suppose
first that we apply the rotation by $\eta$ to an initial state $|\psi\rangle$
about a \emph{known} axis and perform a measurement of the projector
$|\psi\rangle\langle\psi|$ in the rotated state $R_{\mathbf{n}}(\eta)\ket{\psi}$.
The expectation value of the observable $|\psi\rangle\langle\psi|$
is given by 
\begin{equation}
F_{|\psi\rangle}(\eta,\mathbf{n})=|\bra\psi R_{\mathbf{n}}(\eta)|\psi\rangle|^{2}\,,\label{eq: fidelity}
\end{equation}
 i.e.\ by the fidelity between the initial state and the final state.
The fidelity $F_{|\psi\rangle}(\eta,\mathbf{n})$ equals the probability to find the quantum system in the initial state after the rotation. Thus, the probability to detect that the rotation has occurred
is given by the quantity $1-F_{|\psi\rangle}(\eta,\mathbf{n})$. Therefore,
the measurement will be most sensitive if the rotation is applied
to states $|\psi\rangle$ which \emph{minimize} the expression \eqref{eq: fidelity}
for given angle and rotation axis.

Next, suppose that only the rotation angle $\eta$ is well-defined
while the rotation axis is not known, as described in \cite{ChrHer17}.
This situation occurs, for example, when spins prepared in the state $\ket\psi$ are---during the measurement sequence---subjected to a magnetic field whose direction randomly fluctuates on a time scale
much larger than the Larmor period. Measuring the observable $|\psi\rangle\langle\psi|$
on an ensemble of identically prepared systems will now produce a
value of the fidelity \eqref{eq: fidelity} averaged over all possible
spatial directions $\mathbf{n}$. Then, the most suitable quantum
states $\ket\psi$---called \emph{optimal} \emph{quantum rotosensors}
in \cite{ChrHer17}---are determined by the requirement that the
\emph{average fidelity} 
\begin{equation}
{\cal F}_{|\psi\rangle}(\eta)=\frac{1}{4\pi}\int_{\unitS}F_{|\psi\rangle}(\eta,\mathbf{n})\,d\mathbf{n}\,,\label{eq: probability}
\end{equation}
achieve its minimum, for a given value of the parameter $\eta$. 

The fidelity \eqref{eq: fidelity} and its average \eqref{eq: probability} also play a role when setting
up experiments which aim to determine an unknown rotation angle as accurately as possible. This is explained in more detail in Appendix~\ref{Appendix_param}.

For the spin values $j=1/2,1,3/2,2$, optimal quantum rotosensors
have been identified \cite{ChrHer17}, using an approach which combines
analytical and numerical methods. For rotation angles $\eta$ close
to $\pi$, the average fidelity is minimized systematically by \emph{coherent}
spin states. Coherent spin states are strongly localized in phase space and entirely specified by a spatial direction into which they point on the Bloch
sphere \cite{Are72}. For small rotation angles $\eta$, the
average fidelity is minimized by \emph{anticoherent} states, which
are characterized by the fact that they do not manifest any privileged
direction; in this respect, they are as distinct as possible from
coherent states \cite{Zimba06}. The role of anticoherent states
for optimal detection of rotations has also been observed and was subsequently quantified in terms of quantum Fisher information in~\cite{Goldberg18}.
Between these two extreme cases of $\eta\sim0$ and $\eta\sim\pi$, optimal states are neither coherent nor anticoherent
in general. From an experimental point of view, anticoherent and other non-classical spin states have been created 
using a variety of physical systems. For instance, anticoherent states of quantum
light fields have been generated using orbital angular momentum states
of single photons with their usefulness for quantum metrology being
established in~\cite{Bou17}. Non-classical spin states---including
Schrödinger cat states (c.f.\ Sec.~\ref{sec: Optimal-quantum-rotosensors})---of
highly magnetic dysprosium atoms with spin quantum number $j=8$ have
been created in order to enhance the precision of a magnetometer \cite{Cha18}.

The main result of the present paper is a closed-form expression
of the average fidelity ${\cal F}_{|\psi\rangle}(\eta)$, valid for
arbitrary values of $j$. A rather general argument, based solely
on the symmetries of the average fidelity ${\cal F}_{|\psi\rangle}(\eta)$,
shows that it must be a linear combination of the form 
\begin{equation}
{\cal F}_{|\psi\rangle}(\eta)=\varphi_{0}^{(j)}(\eta)+\sum_{t=1}^{\lfloor j\rfloor}\varphi_{t}^{(j)}(\eta)\,\mathcal{A}_{t}(|\psi\rangle),\label{PexpansionAC}
\end{equation}
as explained in detail in Sec.~\ref{sec: Tools and concepts}.
In this expression, the ${\cal A}_{t}(\ket\psi)$ are the anticoherence
measures of a state $\ket\psi$, introduced in \cite{Bag17} and given
explicitly in Eq.~\eqref{ACR}, while the real-valued functions $\varphi_{t}^{(j)}(\eta)$
are trigonometric polynomials independent of $\ket{\psi}$, and $\lfloor j\rfloor$
is the largest integer smaller than or equal to $j$. The main challenge is to
calculate the $\eta$-dependent coefficients $\varphi_{t}^{(j)}(\eta)$,
which we do in Sec.~\ref{sec: Closed-form}.

In earlier works, the average fidelity ${\cal F}_{|\psi\rangle}(\eta)$
had been expressed as a sum of functions of $\eta$ weighted by \emph{state-dependent}
coefficients, upon representing the state in the polarization-tensor
basis \cite{ChrHer17}. The advantage of relation \eqref{PexpansionAC}
is that the average fidelity depends on the state under consideration
only through its measures of anticoherence, and thus it directly relates
to the degree of coherence or anticoherence of the state. Expression
\eqref{PexpansionAC} allows us to identify optimal quantum rotosensors
for spin quantum numbers up to $j=5,$ thereby confirming the role
played by coherent and anticoherent states beyond $j=2$. Readers
mainly interested in the optimal quantum rotosensors may want to directly
consult Sec.~\ref{sec: Optimal-quantum-rotosensors}.

Let us outline the overall argument leading to the expression of the
average fidelity ${\cal F}_{|\psi\rangle}(\eta)$ in \eqref{PexpansionAC}.
In Sec.~\ref{sec: Tools and concepts}, we introduce a number of
tools and concepts feeding into the derivation of \eqref{PexpansionAC}:
first, we discuss the symmetries built into the average fidelity ${\cal F}_{\ket\psi}(\eta)$,
followed by a brief summary of the Majorana representation which enables
us to interpret spin-$j$ states as completely symmetric states of
$N=2j$ qubits. This perspective allows us to introduce, for $1\leqslant t \leqslant \floorj$, the anticoherence
measure $\mathcal{A}_{t}(\ket\psi)$, defined as the linear entropy of the $t$-qubit reduced density matrix of $\ket\psi\bra\psi$. To actually carry
out the integration in Eq.~\eqref{eq: probability}, we will use a
tensor representation (see Sec.~\ref{subsec:Tensor-representation})
of mixed spin-$j$ states generalizing the Bloch representation.
In addition, this representation also enables us to exploit the symmetries
of the average fidelity which can only depend on expressions invariant
under $\mathrm{SU}(2)$ rotations. As shown in Sec.~\ref{subsec:-Invariants},
it is then possible to establish a linear relation between these invariants
and the anticoherence measures $\mathcal{A}_{t}(\ket\psi)$, which finally leads
to \eqref{PexpansionAC}.

Section~\ref{sec: Closed-form} is dedicated to deriving explicit
expressions for the functions $\varphi_{t}^{(j)}(\eta)$. This will
be done in two ways: the first one is based on the fact that anticoherence
measures are explicitly known for certain states, so that the functions
$\varphi_{t}^{(j)}(\eta)$ appear as solutions of a linear system
of equations. The second approach makes use of representations of
the Lorentz group and allows us to obtain a general closed expression.
In Sec.~\ref{sec: Optimal-quantum-rotosensors} we make use of this
closed-form expression to identify the optimal quantum rotosensors.
We conclude with a brief summary given in Sec. \ref{sec:Conclusion}.

\section{Concepts and tools\label{sec: Tools and concepts}}

In this section, we introduce the tools that will be needed to address
the optimality problem described in the Introduction.

\subsection{Notation}

Quantum systems with integer or half-integer spin $j$ are described
by states $\ket{\psi}$ of the Hilbert space $\mathbb{C}^{N+1}$ with $N=2j$,
which carries a $(N+1)$-dimensional representation of the group
SU$(2)$. The components of the angular momentum operator ${\bf J}$
satisfy $[J_{k},J_{\ell}]=i\varepsilon_{k\ell m}J_{m}$, $k,\ell,m\in\{x,y,z\}$, where $\varepsilon_{k\ell m}$ is the Levi-Civita symbol.
Denoting unit vectors in $\mathbb{R}^{3}$ by 
\begin{equation}
\mathbf{n}=\begin{pmatrix}\sin\theta\cos\phi\\
\sin\theta\sin\phi\\
\cos\theta
\end{pmatrix}\,,\quad\theta\in[0,\pi]\,,\quad\phi\in[0,2\pi[\,,\label{eq: unit vector}
\end{equation}
the operator 
\begin{equation}
R_{\mathbf{n}}(\eta)=e^{-i\eta\mathbf{J}\boldsymbol{\cdot}\mathbf{n}}\label{rot}
\end{equation}
describes a rotation by an angle $\eta\in[0,4\pi[$ about the direction
$\mathbf{n}$.

\subsection{Symmetries}

By definition, the average fidelity in \eqref{eq: probability} is
a positive function of the angle $\eta$ and of the state $\ket{\psi}$ and possesses three symmetries:
it is $2\pi$-periodic in $\eta$, symmetric about $\eta=\pi$, and invariant under rotation of $\ket{\psi}$.

Periodicity with period $2\pi$ comes from the fact that $R_{\mathbf{n}}(2\pi)=(-1)^{N}$.
Symmetry about $\eta=\pi$ is equivalent to
\begin{equation}
{\cal F}_{\ket{\psi}}(\eta)={\cal F}_{\ket{\psi}}(2\pi-\eta)\,,\label{eq: reflection symmetry for angle eta}
\end{equation}
which can be shown using $R_{\mathbf{n}}(2\pi-\eta)=(-1)^{N}R_{-\mathbf{n}}(\eta)$
and the fact that the set of directions averaged over in \eqref{eq: probability}
is the same irrespective of the sign of the unit vector ${\bf n}$
since the fidelity \eqref{eq: fidelity} is given by the the \emph{squared}
modulus of the overlap between the states $|\psi\rangle$ and $R_{\mathbf{n}}(\eta)\ket{\psi}$.

Invariance under rotation of $\ket{\psi}$ can be understood in the following way.
Let $R_{\mathbf{m}}(\chi)=e^{-i\chi\mathbf{J}\boldsymbol{\cdot}\mathbf{m}}$
be a unitary operator
representing a rotation in $\mathbb{R}^{3}$ by an angle $\chi\in[0,4\pi[$
about the direction $\mathbf{m}$, acting on a state $|\psi\rangle\in\mathbb{C}^{N+1}$.
Then the average fidelities ${\cal F}$ associated with the states $|\psi\rangle$
and $|\psi^R\rangle\equiv R_{\mathbf{m}}(\chi)|\psi\rangle$ are equal. Indeed,  
we have
\begin{equation}
F_{|\psi^R\rangle}(\eta,\mathbf{n})=\bra{\psi}R_{\mathbf{m}}(\chi)^{\dagger}R_{\mathbf{n}}(\eta)R_{\mathbf{m}}(\chi)|\psi\rangle\label{eq: trf of F under rotations}
\end{equation}
and
\begin{align}
R_{\mathbf{m}}(\chi)^{\dagger}R_{\mathbf{n}}(\eta)R_{\mathbf{m}}(\chi) & =e^{-i\eta (R_{\mathbf{m}}(\chi)^{\dagger}\mathbf{J}R_{\mathbf{m}}(\chi))\boldsymbol{\cdot}\mathbf{n}}\nonumber \\
 & =e^{-i\eta(R\mathbf{J})\boldsymbol{\cdot}\mathbf{n}}=e^{-i\eta\mathbf{J}\boldsymbol{\cdot}\mathbf{n}^R}\,,\label{eq: trf of R under rotation}
\end{align}
with ${\bf n}^R\equiv R^{T}\mathbf{n}$ the vector obtained by the rotation $R\in$ SO$(3)$ associated with $R_{\mathbf{m}}(\chi)$.
Due to the invariance under rotations of the unit-ball region $\unitS$
appearing in \eqref{eq: probability} (invariance of
the Haar measure used), the result of the integration will be the
same, leading to 
\begin{align}
{\cal F}_{|\psi^R\rangle}(\eta) & =\frac{1}{4\pi}\int_{\unitS}F_{|\psi^R\rangle}(\eta,\mathbf{n})\,d\mathbf{n}\nonumber \\
 & =\frac{1}{4\pi}\int_{\unitS}F_{|\psi\rangle}(\eta,\mathbf{n})\,d\mathbf{n}={\cal F}_{|\psi\rangle}(\eta)\,.\label{eq: invariance of av fid}
\end{align}
This invariance of the fidelity can be seen in a geometrically appealing way by use of the Majorana representation, 
which we consider now.

\subsection{Majorana representation of pure spin states \label{subsec:Majorana-representation}}

The Majorana representation establishes a one-to-one correspondence
between spin-$j$ states and $N=2j$-qubit states that
are invariant under permutation of their constituent qubits (see e.g.~\cite{Bie81,Zyczkowski_book,Coe98}). It allows
to geometrically visualise a pure spin-$j$ state as $N$ points on
the unit sphere associated with the Bloch vectors of the $N$ qubits.
The Majorana points are often referred to as stars, and the 
whole set of Majorana points of a given state as its Majorana constellation.
Considering a spin-$j$ state $\ket{\psi}$ as an $N$-qubit state,
any local unitary (LU) operation $U=u^{\otimes N}$ with $u\in \mathrm{SU}(2)$
transforms $\ket{\psi}$ into a state whose Majorana
constellation is obtained by the constellation of $\ket{\psi}$ rotated
by the SO$(3)$ rotation associated with $u$. Spin-coherent states take a very simple form in the Majorana representation, as they can be seen as  the tensor product $\ket{\phi}^{\otimes N}$ of some spin-$1/2$ state $\ket{\phi}$. Their constellation thus reduces to an $N$-fold degenerate point.

The fidelity \eqref{eq: fidelity} is given by the squared modulus of the overlap between $|\psi\rangle$
and $R_{\mathbf{n}}(\eta)\ket{\psi}$. Since the Majorana constellation of 
$R_{\mathbf{n}}(\eta)\ket{\psi}$ is obtained by rigidly rotating that of $|\psi\rangle$, the fidelity \eqref{eq: fidelity} only depends on the relative positions of
these two sets of points. The \emph{average} transition probability
${\cal F}_{|\psi\rangle}(\eta)$ is obtained by integrating over all possible constellations obtained
by rigid rotations of the Majorana constellation of 
$|\psi\rangle$, and therefore it must be invariant under LU.
In other words, the equality \eqref{eq: invariance of av fid}
takes the form ${\cal F}_{|\psi\rangle}(\eta)={\cal F}_{u^{\otimes N}|\psi\rangle}(\eta)$.

\subsection{Anticoherence measures \label{subsec:Anticoherence-measures}}

An order-$t$ \emph{anticoherent} state $\ket\chi$
is defined by the property that $\langle\chi|(\mathbf{J}\boldsymbol{\cdot}\mathbf{n})^{k}|\chi\rangle$
is independent of the vector $\mathbf{n}$ for all $k=1,\ldots,t$. In the Majorana representation, it is characterized by the fact that its $t$-qubit reduced density matrix is the maximally mixed state in the symmetric sector~\cite{prl}.

The degree of coherence or $t$-anticoherence of a spin-$j$ pure
state $|\psi\rangle$ can be measured by the quantities $\mathcal{A}_{t}(|\psi\rangle)$,
which are positive-valued functions of $|\psi\rangle$ \cite{Bag17}.
Let $\rho_{t}=\tr_{\neg t}\left[|\psi\rangle\langle\psi|\right]$ be
the $t$-qubit reduced density matrix of the state $|\psi\rangle$
interpreted as a $N$-qubit symmetric state with $N=2j$; it is obtained by taking
the partial trace over all but $t$ qubits (it does not matter which
qubits are traced over since $|\psi\rangle$ is a symmetric state).
The measures $\mathcal{A}_{t}(|\psi\rangle)$ are defined as the rescaled
linear entropies 
\begin{equation}
\mathcal{A}_{t}(|\psi\rangle)=\frac{t+1}{t}\left(1-\tr\left[\rho_{t}^{2}\right]\right)\,,\label{ACR}
\end{equation}
where $\tr\left[\rho_{t}^{2}\right]$ is the purity of $\rho_{t}$.
Thus, anticoherence measures are quartic in the state $|\psi\rangle$
and range from $0$ to $1$, and are invariant
under SU$(2)$ rotations. Spin-coherent states are characterized by
pure reduced states and thus are the only states such that $\mathcal{A}_{t}=0$. Anticoherent
states to order $t$ are characterized by $\rho_t=\mathbb{1}/(t+1)$ and thus are the only states such that $\mathcal{A}_{t}=1$. In particular, if a state $|\psi\rangle$
is anticoherent to some order $t$,
then it is necessarily anticoherent to all lower orders $t'=1,\ldots,t$ since reductions of the maximally mixed state are maximally mixed.

While for any state we have $0\leqslant\mathcal{A}_{t}\leqslant1$,
not all possible tuples $(\mathcal{A}_{1},\mathcal{A}_{2},\ldots)$
are realised by a physical state $|\psi\rangle$. For instance, since
$\mathcal{A}_{t}=1$ implies that $\mathcal{A}_{t'}=1$ for all $t'\leqslant t$,
the choice $\mathcal{A}_{2}=1$ and $\mathcal{A}_{1}<1$ cannot correspond
to any state. We denote the domain of admissible values of the measures
$\mathcal{A}_{t}$ by $\Omega$.

\subsection{Tensor representation of mixed states \label{subsec:Tensor-representation}}

We now introduce a tensor representation of an arbitrary (possibly
mixed) spin-$j$ state $\rho$ acting on a $(N+1)$-dimensional Hilbert
space with $N=2j$, following~\cite{prl}. Any state can be expanded as
\begin{equation}
\rho=\frac{1}{2^{N}}\,x_{\mu_{1}\mu_{2}\ldots\mu_{N}}S_{\mu_{1}\mu_{2}\ldots\mu_{N}}.\label{rhoarbitrary}
\end{equation}
Here and in what follows, we use Einstein summation convention
for repeated indices, with Greek indices running from $0$ to $3$ and Latin indices running from $1$ to $3$.
Here, the $S_{\mu_{1}\mu_{2}\ldots\mu_{N}}$ are $(N+1)\times(N+1)$
Hermitian matrices invariant under permutation of the indices. 

The $x_{\mu_{1}\mu_{2}\ldots\mu_{N}}$ are real coefficients also invariant
under permutation of their indices, which enjoy what we call the tracelessness
property 
\begin{equation}
\sum_{a=1}^{3}x_{aa\mu_{3}\ldots\mu_{N}}=x_{00\mu_{3}\ldots\mu_{N}}\,,\quad\forall\;\mu_{3},\ldots,\mu_{N}.\label{traceless}
\end{equation}
Whenever $x_{\mu_{1}\mu_{2}\ldots\mu_{N}}$ has some indices equal
to $0$, we take the liberty to omit them, so that e.g.~for a spin-$3$
state $x_{110200}$ may be written $x_{112}$ (recall that the order
of the indices does not matter). In the case of a spin-coherent state
given by its unit Bloch vector $\mathbf{n}=(n_{1},n_{2},n_{3})$, the coefficients
in \eqref{rhoarbitrary} are simply given by $x_{\mu_{1}\mu_{2}\ldots\mu_{N}}=n_{\mu_{1}}n_{\mu_{2}}\ldots n_{\mu_{N}}$,
with $n_{0}=1$.

In the following, we will make use of two essential properties of
the tensor representation. Namely, let us consider a state $\rho$
with coordinates $x_{\mu_{1}\mu_{2}\ldots\mu_{N}}$ in the expansion
\eqref{rhoarbitrary}. Then, the tensor coordinates of the $t$-qubit
reduced state $\rho_{t}$ in the expansion \eqref{rhoarbitrary} are
simply given by $x_{\mu_{1}\mu_{2}\ldots\mu_{t}}=x_{\mu_{1}\mu_{2}\ldots\mu_{t}0...0}$.
Thus, since we omit the zeros in the string $\mu_{1}\mu_{2}\ldots\mu_{N}$,
the tensor coordinates of $\rho_{t}$ and $\rho$ coincide for any
string of $k\leq t$ nonzero indices.

The second property we use is that for states $\rho$ and $\rho'$
in the form \eqref{rhoarbitrary} with tensor coordinates respectively
$x_{\mu_{1}\mu_{2}\ldots\mu_{N}}$ and $x'_{\mu_{1}\mu_{2}\ldots\mu_{N}}$
we have 
\begin{equation}
\tr\left[\rho\rho'\right]=\frac{1}{2^{N}}\sum_{\mu_{1},\mu_{2},...,\mu_{N}}x_{\mu_{1}\mu_{2}\ldots\mu_{N}}x'_{\mu_{1}\mu_{2}\ldots\mu_{N}}.
\end{equation}
Note that this equality holds despite the fact that the $S_{\mu_{1}\mu_{2}\ldots\mu_{N}}$ are not orthogonal; this property follows from the fact that these matrices form a $2^N$-tight frame, see~\cite{prl}.
In particular, for a pure state $\rho=\ket{\psi}\bra{\psi}$, the
equality $\tr\rho^{2}=1$ translates into 
\begin{equation}
\sum_{\mu_{1},\mu_{2},...,\mu_{N}}x_{\mu_{1}\mu_{2}\ldots\mu_{N}}^{2}=2^{N},\label{forpure}
\end{equation}
while the purity of the reduced density matrix $\rho_{t}$ reads 
\begin{equation}
\tr\left[\rho_{t}^{2}\right]=\frac{1}{2^{t}}\sum_{\mu_{1},\mu_{2},...,\mu_{t}}x_{\mu_{1}\mu_{2}\ldots\mu_{t}}^{2}\,.\label{trrt2}
\end{equation}
The normalization condition $\tr\left[\rho\right]=1$ imposes $x_{00\ldots0}=1$.
A consequence of \eqref{traceless} is then that $\sum_{a=1}^3x_{aa}=1$.

\subsection{SU$(2)$-Invariants \label{subsec:-Invariants}}

If $u\in$ SU(2) and $R\in$ SO(3) is the corresponding rotation matrix,
then the tensor coordinates of $U\rho U^{\dagger}$ with $U=u^{\otimes N}$
are the $\mathsf{R}_{\mu_{1}\nu_{1}}\ldots\mathsf{R}_{\mu_{N}\nu_{N}}x_{\nu_{1}\ldots\nu_{N}}$
where $\mathsf{R}$ is the $4\times4$ orthogonal matrix 
\begin{equation}
\mathsf{R}=\left(\begin{array}{c|c}
\begin{array}{c}
1\end{array} & \begin{array}{c}
0\end{array}\\
\hline \begin{array}{c}
0\end{array} & \begin{array}{c}
R\end{array}
\end{array}\right).\label{matrixR}
\end{equation}
That is, $x_{\mu_{1}\mu_{2}\ldots\mu_{N}}$ transforms as a tensor.
Under such transformations, $x_{\mu}x_{\mu}$ goes into $\mathsf{R}_{\mu\nu}\mathsf{R}_{\mu\nu'}x_{\nu}x_{\nu'}=(\mathsf{R}^{T}\mathsf{R})_{\nu'\nu}x_{\nu}x_{\nu'}=x_{\nu}x_{\nu}$,
where the last equality comes from orthogonality of $\mathsf{R}$.
Thus $x_{\mu}x_{\mu}$ is an SU(2) invariant. Similarly, $x_{\mu}x_{\mu\nu}x_{\nu}$
and, more generally, any product of the $x_{\mu_{1}\mu_{2}\ldots\mu_{N}}$
such that all indices are contracted (i.e.\ summed from $0$ to $3$),
are invariant under SU(2) action on $\rho$. One can then show by
induction that products of terms $x_{a_{1}a_{2}\ldots a_{k}}$ with
$k\leqslant N$ where all indices appear in pairs and are summed from
$1$ to $3$ are also SU(2) invariant. For instance, $x_{a}x_{a}$,
$x_{ab}x_{ab}$, $x_{ab}x_{bc}x_{ca}$, $x_{a}x_{ab}x_{b}$ are such
invariants.

Invariants of degree 1 in $x$ are of the form $x_{a_{1}a_{2}\ldots a_{2k}}$,
where the $a_{i}$ appear in pairs. Since the order of indices is
not relevant, these invariants are in fact of the form $x_{a_{1}a_{1}a_{2}a_{2}\ldots a_{k}a_{k}}$.
Because of Eq.~\eqref{traceless}, each pair can be replaced by zeros
in the string, so that $x_{a_{1}a_{1}a_{2}a_{2}\ldots a_{k}a_{k}}=x_{00\ldots0}=1$.
Therefore, there is no invariant of degree 1. The invariants of degree
2 are products of the form $x_{a_{1}a_{2}\ldots a_{k}}x_{b_{1}b_{2}\ldots b_{k'}}$
where indices appear in pairs and are summed from $1$ to $3$. If
the two indices of a pair appear in the same index string ($a_{1}a_{2}\ldots a_{k}$
or $b_{1}b_{2}\ldots b_{k'}$), then from Eq.~\eqref{traceless},
they can again be replaced by zeros and discarded. Thus the invariants
of degree 2 are $\kappa_{1}=x_{a}x_{a}$, $\kappa_{2}=x_{ab}x_{ab}$,
and more generally, for $1\leqslant r\leqslant N$, 
\begin{equation}
\kappa_{r}=x_{a_{1}a_{2}...a_{r}}x_{a_{1}a_{2}...a_{r}}.\label{defkappa}
\end{equation}

Using \eqref{ACR} and \eqref{trrt2} one can express the invariants
$\kappa_{r}$ in terms of a linear combination of the $\mathcal{A}_{t}$.
Indeed, grouping together terms with the same number of nonzero indices
in \eqref{trrt2} yields 
\begin{equation}
\tr\left[\rho_{t}^{2}\right]=\frac{1}{2^{t}}\sum_{\mu_{1},\mu_{2},...,\mu_{t}}x_{\mu_{1}\mu_{2}\ldots\mu_{t}}^{2}=\frac{1}{2^{t}}\sum_{r=0}^{t}\binom{t}{r}\kappa_{r}\,.\label{trbis}
\end{equation}
Inverting that relation via the binomial inversion formula, we obtain
\begin{equation}
\kappa_{r}=\sum_{t=0}^{r}(-1)^{t+r}\,2^{t}\binom{r}{t}\tr\left[\rho_{t}^{2}\right]\,,\label{invrel}
\end{equation}
and by use of \eqref{ACR} we finally can express the $\mathrm{SU}(2)$-invariants
in terms of anticoherence measures, 
\begin{equation}
\kappa_{r}=\sum_{t=0}^{r}(-1)^{t+r}\,2^{t}\binom{r}{t}\left(1-\frac{t}{t+1}\mathcal{A}_{t}\right)\label{kappar}
\end{equation}
for $r=1,\ldots,N$.

\subsection{General form of the average fidelity}

Let us now explain why the average fidelity ${\cal F}_{|\psi\rangle}(\eta)$
given in Eq.~\eqref{PexpansionAC} is a linear combination of the
lowest $\left\lfloor j\right\rfloor $ anticoherent measures $\mathcal{A}_{t}$.
Due to its rotational symmetry, the average fidelity ${\cal F}_{|\psi\rangle}(\eta)$---when
considered as a function of the tensor coordinates $x_{\mu_{1}\mu_{2}\ldots\mu_{N}}$---can
only involve invariants constructed from these coordinates. With ${\cal F}_{|\psi\rangle}(\eta)$
being quadratic in $\rho=\ket{\psi}\bra{\psi}$, it must also be quadratic
in $x$. As there is no invariant of degree 1, the only invariants
that can appear in the expression of ${\cal F}_{|\psi\rangle}(\eta)$
are the invariants $\kappa_{r}$ defined in \eqref{defkappa}. Since
the quantity ${\cal F}_{|\psi\rangle}(\eta)$ is quadratic it must
be a linear combination of the coefficients $\kappa_{r}$ which, according
to Eq.~\eqref{kappar}, implies that ${\cal F}_{|\psi\rangle}(\eta)$
is also a linear combination of the $\mathcal{A}_{t}$. Furthermore,
the identity 
\begin{equation}
\tr\left[\rho_{t}^{2}\right]=\tr\left[\rho_{N-t}^{2}\right]\,,\label{eq: t vs N-t symmetry}
\end{equation}
which holds for any pure state, means that the anticoherence measures
$\mathcal{A}_{t}$ for $t>N/2$ can be expressed in terms of the measures
$\mathcal{A}_{t}$ for $t< N/2$. Therefore, \eqref{PexpansionAC}
is the most general form the fidelity ${\cal F}_{|\psi\rangle}(\eta)$
can take, with the dependence in $\eta$ being only in the coefficients
of the measures $\mathcal{A}_{t}$.

\subsection{Generalizations}
It is worth stressing that the form \eqref{PexpansionAC} for the average fidelity also holds for more general types of average fidelity
\begin{equation}\label{genfid}
\frac{1}{4\pi}\int_{S^{2}}|\langle\psi|U_{\mathbf{n}}(\eta)|\psi\rangle|^{2}\,d\mathbf{n}
\end{equation}
between a state $|\psi\rangle$ and its image under the unitary
\begin{equation}
U_{\mathbf{n}}(\eta)=e^{-i\eta\, f(\mathbf{J}\boldsymbol{\cdot}\mathbf{n})},
\end{equation}
where $f$ is an arbitrary real analytic function, ensuring that $f(\mathbf{J}\boldsymbol{\cdot}\mathbf{n})$ is an Hermitian operator. Indeed, from an argument similar to that of Sec.~\ref{subsec:-Invariants}, the generalized fidelity \eqref{genfid} can be expressed as a function of the $\kappa_r$ and hence of the $\mathcal{A}_t$. An interesting case is when $U_{\mathbf{n}}(\eta)$ is a spin-squeezing operator, which corresponds to choosing $f(\mathbf{J}\boldsymbol{\cdot}\mathbf{n})=(\mathbf{J}\boldsymbol{\cdot}\mathbf{n})^2$.
Moreover, if we now consider the quantities
\begin{equation}\label{genfid2}
\frac{1}{4\pi}\int_{S^{2}}|\langle\psi|U_{\mathbf{n}}(\eta)|\psi\rangle|^{2k}\,d\mathbf{n}
\end{equation}
with integer $k\geqslant 2$, the same arguments show that they are linear combinations of higher-order invariants, leading to generalizations of the relation \eqref{kappar}.

\section{Closed form of the average fidelity \label{sec: Closed-form}}

In this section we derive the angular functions $\varphi_{t}^{(j)}(\eta)$,
which characterize the fidelity through \eqref{PexpansionAC}, in
two different ways. The first method (subsection \ref{closed1})
is based on the fact that anticoherence measures can be evaluated explicitly
for Dicke states. The second method (subsection \ref{closed2}) exploits
a tensor representation of spin states \cite{prl} which uses Feynman
rules from relativistic spin theory. These approaches are independent and we checked, for all integers and half-integers $j$ up to $26$, that as expected they yield the same angular functions. Technical detail is delegated to appendices in both cases.

\subsection{Derivation based on anticoherence measures for Dicke states \label{closed1} }

In the following, we will work in the standard angular momentum basis
of $\mathbb{C}^{N+1}$, for positive integer or half-integer value of $j=N/2$. It consists of the
Dicke states $\left\{ \ket{j,m},\left|m\right|\leq j\right\} $ given
by the common eigenstates of $\mathbf{J}^{2}$, the square of the
angular momentum operator $\mathbf{J}$, and of its $z$-component
$J_{z}$. In this basis, any spin-$j$ state $|\psi\rangle$ can
be expanded as 
\begin{equation}
|\psi\rangle=\sum_{m=-j}^{j}c_{m}\,\ket{j,m}\,,\label{eq:jm_decomp}
\end{equation}
with $c_{m}\in\mathbb{C}$ and $\sum_{m=-j}^{j}|c_{m}|^{2}=1$. 

The first derivation is based on the fact that both the measures of
$t$-anticoherence $\mathcal{A}_{t}(|j,m\rangle)$ and the average
fidelities ${\cal F}_{|j,m\rangle}(\eta)$ can be determined explicitly
for Dicke states. Their measures of $t$-anticoherence are given by
\begin{equation}
\mathcal{A}_{t}(|j,m\rangle)=\frac{t+1}{t}\left[1-\frac{\sum_{\ell=0}^{t}\binom{j+m}{t-\ell}^{2}\binom{j-m}{j-m-\ell}^{2}}{\binom{2j}{t}^{2}}\right].
\label{ACRDicke}
\end{equation}
They can readily be obtained from the purities $\tr\left[\rho_{t}^{2}\right]$ for a state of the form \eqref{eq:jm_decomp}, which were calculated in \cite{Bag17} in terms of the coefficients $c_{m}$ and read
\begin{equation}
\tr\left[\rho_{t}^{2}\right]=\sum_{q,\ell=0}^t\left| \sum_{k=0}^{2j-t} c_{j-k-\ell}^* \, c_{j-k-q} \,\Gamma_k^{\ell q} \right|^2
\label{puritiescm}
\end{equation}
with
\begin{equation}
\Gamma_k^{\ell q}=\frac{\sqrt{\binom{2j-k-q}{t-q}\binom{2j-k-\ell}{t-\ell}\binom{k+q}{k}\binom{k+\ell}{k}}}{\binom{2j}{t}}.
\end{equation}
As for the
fidelity, the calculation is done in Appendix \ref{sec: appendix C (Dicke)}
and yields 
\begin{equation}
{\cal F}_{|j,m\rangle}(\eta)=\frac{1}{(2j+1)^{2}}\sum_{\ell=0}^{2j}(2\ell+1)(C_{jm\ell0}^{jm}\,\chi_{\ell}^{j}(\eta))^{2}\,,\label{PDicke}
\end{equation}
with Clebsch-Gordan coefficients $C_{jm\ell0}^{jm}$ and the functions
$\chi_{\ell}^{j}(\eta)$ defined in Eqs.~\eqref{chilj}--\eqref{chij}.
The angular functions $\varphi_{t}^{(j)}(\eta)$ are then solutions
of the system of linear equations 
\begin{equation}
\left\{ \begin{array}{l}
{\cal F}_{|j,m\rangle}(\eta)=\varphi_{0}^{(j)}(\eta)+\sum_{t=1}^{\lfloor j\rfloor}\varphi_{t}^{(j)}(\eta)\,\mathcal{A}_{t}(|j,m\rangle)\\[8pt]
\mathrm{for}\;\,m=j,j-1,\ldots,j-\lfloor j\rfloor.
\end{array}\right.\label{syseq}
\end{equation}
This system can easily be solved for the lowest values of $j$. A
general (but formal) solution can then be obtained by inverting the
system \eqref{syseq}. 

\subsection{Derivation based on relativistic Feynman rules and tensor representation
of spin states \label{closed2}}

The second approach allows us to derive a closed-form expression for
the functions $\varphi_{t}^{(j)}(\eta)$. It is based on an expansion
of the operator 
\begin{equation}
\Pi^{(j)}(q)\equiv(q_{0}^{2}-|\bq|^{2})^{j}\,e^{-2\theta_{q}\,\hbq\boldsymbol{\cdot}\bJ},\label{lorentzboost}
\end{equation}
with $\tanh\theta_{q}=-|\bq|/q_{0}$ and $\hbq=\bq/|\bq|$, as a multivariate
polynomial in the variables $q_{0},q_{1} ,q_{2},q_{3}$. This operator is a
$(N+1)$-dimensional representation (with $N=2j$) of a Lorentz boost in the direction
of the 4-vector $q=(q_{0},\bq)=(q_{0},q_{1},q_{2},q_{3})$. As shown in \cite{Wei64},
it can be written as 
\begin{equation}
\Pi^{(j)}(q)=(-1)^{2j}q_{\mu_{1}}q_{\mu_{2}}\ldots q_{\mu_{2j}}S_{\mu_{1}\mu_{2}\ldots\mu_{2j}}.\label{egaliteweinberg}
\end{equation}
The identification of Eqs.~\eqref{lorentzboost} and \eqref{egaliteweinberg}
defines the $(N+1)\times(N+1)$ matrices $S_{\mu_{1}\ldots\mu_{N}}$
appearing in \eqref{rhoarbitrary} (see~\cite{prl} for detail).
Taking 
\begin{equation}
q_{0}=i\cot(\eta/2)\quad\mbox{and}\quad q_{i}=n_{i}\,,\quad i=1,2,3\,,\label{eq: q0 + qi}
\end{equation}
in \eqref{lorentzboost}, we see that $\Pi^{(j)}(q)$ reduces to a
rotation operator, 
\begin{equation}
R_{\mathbf{n}}(\eta)=e^{-i\eta\mathbf{J}\boldsymbol{\cdot}\mathbf{n}}=\frac{\Pi^{(j)}(q)}{m^{N}}\label{eq: rot as Pi}
\end{equation}
with 
\begin{equation}
m^{2}=q_{0}^{2}-|\bq|^{2}=-\frac{1}{\sin^{2}(\eta/2)}.\label{eqm}
\end{equation}
Moreover, for a state $\rho$ given by \eqref{rhoarbitrary} we have
\begin{equation}
\tr\left[\rho\,\Pi^{(j)}(q)\right]=(-1)^{N}x_{\mu_{1}\mu_{2}\ldots\mu_{N}}q_{\mu_{1}}\ldots q_{\mu_{N}},\label{eq:xq-1}
\end{equation}
according to Eq.~(24) of \cite{prl}, which holds for any 4-vector
$q$. Thus, with $\rho=\ket{\psi}\bra{\psi}$, using the identity
\eqref{eq: rot as Pi} and the expansion \eqref{egaliteweinberg}
for the rotation operator in \eqref{eq: fidelity} allows us to
explicitly perform the integral in Eq.~\eqref{eq: probability}, resulting
in 
\begin{equation}
\begin{aligned} 
{\cal F}_{|\psi\rangle}(\eta) ={}& \frac{1}{4\pi}\int_{\unitS}|\bra{\psi}R_{\mathbf{n}}(\eta)\ket{\psi}|^{2}\,d\mathbf{n}\\
={}& \frac{1}{4\pi}\int_{\unitS}\left|\textrm{tr}\left[\rho\frac{\Pi^{(j)}(q)}{m^{N}}\right]\right|^{2}d\mathbf{n}\\
={}& (-1)^{N}\frac{x_{\mu_{1}\ldots\mu_{N}}x_{\nu_{1}\ldots\nu_{N}}}{4\pi}\\
& \times \int_{\unitS}\frac{q_{\mu_{1}}\ldots q_{\mu_{N}}q_{\nu_{1}}^{*}\ldots q_{\nu_{N}}^{*}}{m^{2N}}\,d\mathbf{n},
\end{aligned}
\label{integxx-1}
\end{equation}
where $*$ denotes complex
conjugation (which acts on $q_{0}$ only because of the choice \eqref{eq: q0 + qi}
and using $|m|^{2}=-m^{2}$). Each term $q_{\mu_{1}}\ldots q_{\nu_{N}}^{*}$
with $2(N-k)$ indices equal to 0 is proportional to 
\begin{equation}
\frac{q_{0}^{2(N-k)}}{m^{2N}}=(-1)^{k}\sin^{2k}\left(\frac{\eta}{2}\right)\cos^{2(N-k)}\left(\frac{\eta}{2}\right).\label{q0k-1}
\end{equation}
For the remaining $2k$ nonzero indices, we have from \eqref{eq: q0 + qi}
that $q_{i}=n_{i}$, so that \eqref{integxx-1} involves an integral
of the form 
\begin{equation}
\frac{1}{4\pi}\int_{\unitS}n_{a_{1}}n_{a_{2}}\ldots n_{a_{2k}}\,d\mathbf{n}\,,\qquad1\leqslant a_{i}\leqslant3 \,.\label{ints-1}
\end{equation}
These integrals
are performed in Appendix \ref{appexplicit}. The integrals \eqref{ints-1} are in fact precisely
given by the tensor coordinates $x_{a_{1}a_{2}\ldots a_{2k}}^{(0)}$
of the maximally mixed state, whose expression is
explicitly known. One can therefore rewrite \eqref{integxx-1} as
\begin{equation}
\begin{aligned} & {\cal F}_{|\psi\rangle}(\eta)=\sum_{k=0}^{N}(-1)^{N}\frac{q_{0}^{2(N-k)}}{m^{2N}}\\
 & \times\hspace{-0.5cm}\sum_{\genfrac{}{}{0pt}{1}{\boldmu{,}\bn}{2(N-k)\textrm{zeros}}{}}\hspace{-0.5cm}(-1)^{\textrm{nr of 0 in }\bn}x_{\mu_{1}\ldots\mu_{N}\nu_{1}\ldots\nu_{N}}^{(0)}x_{\mu_{1}\ldots\mu_{N}}x_{\nu_{1}\ldots\nu_{N}}\,,
\end{aligned}
\label{Ptot-1}
\end{equation}
where the sum over $\boldmu{,}\bn$ runs over all strings of indices
(between 0 and 3) containing $2(N-k)$ zeros. An explicit expression
for this sum is derived in Appendix~\ref{appexplicit}, leading to
the compact expression
\begin{equation}
{\cal F}_{|\psi\rangle}(\eta)=\sum_{k=0}^{N}\sin^{2k}\left(\frac{\eta}{2}\right)\cos^{2(N-k)}\left(\frac{\eta}{2}\right)\sum_{t=0}^{N}a_{t,k}^{(j)}\;\tr\left[\rho_{t}^{2}\right]\,,\label{Ptotmain}
\end{equation}
with numbers 
\begin{equation}
a_{t,k}^{(j)}=\frac{4^{t}(-1)^{k+t}\binom{2N}{2k}\binom{k}{t}\binom{2N-2t}{N-t}}{(2k+1)\binom{2N}{N}}.\label{ajtk}
\end{equation}
Note that the sum over $k$ in \eqref{Ptotmain} can start at $k=t$ because
the factor $\binom{k}{t}$ in $a_{t,k}^{(j)}$ implies that $a_{t,k}^{(j)}=0$
for $t>k$. Using the symmetry $\tr\left[\rho_{t}^{2}\right]=\tr\left[\rho_{N-t}^{2}\right]$ we may rewrite \eqref{Ptotmain}
as 
\begin{equation}
\begin{aligned}{\cal F}_{|\psi\rangle}(\eta)={} & \sum_{k=t}^{N}\sin^{2k}\left(\frac{\eta}{2}\right)\cos^{2(N-k)}\left(\frac{\eta}{2}\right)\\
 & \times\sum_{t=0}^{\lfloor j\rfloor}\left(a_{t,k}^{(j)}+a_{N-t,k}^{(j)}\right)\left(1-\frac{\delta_{jt}}{2}\right)\tr\left[\rho_{t}^{2}\right].
\end{aligned}
\label{Ptotmainsym}
\end{equation}
From \eqref{ACR} we obtain a relation between $\mathcal{A}_{t}$
and $\tr\left[\rho_{t}^{2}\right]$, namely $\tr\left[\rho_{t}^{2}\right]=1-\frac{t}{t+1}\mathcal{A}_{t}$, which
yields the explicit expression of the polynomials $\varphi_{t}^{(j)}(\eta)$
in Eq.~\eqref{PexpansionAC} as 
\begin{equation}
\varphi_{t}^{(j)}(\eta)=\sum_{k=t}^{N}b_{t,k}^{(j)}\,\sin^{2k}\left(\frac{\eta}{2}\right)\cos^{2(N-k)}\left(\frac{\eta}{2}\right),\label{Phimain}
\end{equation}
with coefficients 
\begin{equation}
b_{t,k}^{(j)}=\left\{ \begin{array}{ll}
{\displaystyle -\frac{t}{t+1}\left(a_{t,k}^{(j)}+a_{N-t,k}^{(j)}\right)\left(1-\frac{\delta_{jt}}{2}\right)} & t\neq0\\
{\displaystyle \frac{\binom{N}{k}}{2k+1}} & t=0\,.
\end{array}\right.\label{btk}
\end{equation}
Note that although $q_{0}$ and $m$ are not well-defined for $\eta=0$,
the ratio in \eqref{q0k-1} always is, so that the expression above
is valid over the whole range of values of $\eta$. For spin-coherent
states, all $\mathcal{A}_{t}$ vanish and thus ${\cal F}_{|\psi\rangle}(\eta)=\varphi_{0}^{(j)}(\eta)$
from Eq.~\eqref{PexpansionAC}, which coincides with the expression
obtained in~\cite{ChrHer17}. For the smallest values of $j$, we
recover the functions obtained in Section \ref{closed1}. In the following
section, we will use the functions $\varphi_{t}^{(j)}(\eta)$ given
in \eqref{Phimain} to identify optimal quantum rotosensors.

\section{Optimal quantum rotosensors \label{sec: Optimal-quantum-rotosensors}}

\subsection{Preliminary remarks}

We now address the question of finding the states $|\psi\rangle$
which minimize the average fidelity ${\cal F}_{|\psi\rangle}(\eta)$
for fixed rotation angles $\eta$. According to Eq.~\eqref{PexpansionAC}, the fidelity is a \emph{linear} function of the anticoherence measures
$\mathcal{A}_{t}$ with $1\leqslant t\leqslant\lfloor j\rfloor$. Linearity, when combined with the fact that the domain $\Omega$, over which the measures $\mathcal{A}_{t}$ vary, is \emph{bounded} implies that the fidelity must attain its minimum on the boundary. The minimization
problem thus amounts to characterizing this domain $\Omega$. Unfortunately,
even for the smallest values of $j$, no simple descriptions of this
domain are known.

We will first determine the states minimizing the $2\pi$-periodic
average fidelity for values of $j$ up to $j=7/2$, with the rotation
angle taking values in the interval $\eta\in[0,\pi]$ (which is sufficient
due to the symmetry \eqref{eq: reflection symmetry for angle eta}).
Then we will examine the limiting case of angles $\eta$ close to
$0$ for arbitrary values of the quantum number $j$. Throughout this
section, we will expand arbitrary states with spin $j$ in terms of
the Dicke states, as shown in Eq.~\eqref{eq:jm_decomp}. 

For spins up to $j=2$ the states minimizing the average fidelity
${\cal F}_{\ket\psi}(\eta)$ are known \cite{ChrHer17}. In Sec.~\ref{subsec:jupto2},
we show that our approach based on the expression \eqref{PexpansionAC}
correctly reproduces these results. Then, in Sec.~\ref{subsec:ju5o2pto7o2},
we consider the minimization problem for spin quantum numbers up to
$j=7/2$, mainly identifying the optimal rotosensors within various
ranges of the rotation angle $\eta$ by numerical techniques. More
specifically, for a fixed angle $\eta$, ${\cal F}_{\ket\psi}(\eta)$
is a function of the $\mathcal{A}_{t}$ which can be parametrized
by the complex coefficients $c_{m}$ entering the expansion \eqref{eq:jm_decomp}
of the state $|\psi\rangle$ in the Dicke basis (see Eq.~\eqref{puritiescm}). We search numerically for the minimum value of ${\cal F}_{\ket\psi}(\eta)$
with respect to the $c_{m}$, taking into account the normalization
condition $\sum_{m}|c_{m}|^{2}=1$. In most cases this numerical search
converges towards states which have simple analytic expressions
which are the ones that we give. For each value of $j$, we performed this search at about 1000 evenly spaced values of $\eta$ in order to explore the whole range of rotation angles. Whenever we find a region of values of $\eta$ in which $|\psi_{1}\rangle$ is the optimal state adjacent to a region where $|\psi_{2}\rangle$ is optimal, at the critical angle separating these  two regions, one should have ${\cal F}_{|\psi_{1}\rangle}(\eta)={\cal F}_{|\psi_{2}\rangle}(\eta)$ because the average fidelity ${\cal F}_{|\psi\rangle}(\eta)$ is a continuous function of $|\psi\rangle$. Therefore, the critical angle is a solution of the equation 
\begin{equation}\label{eqcritical}
\sum_{t=1}^{\lfloor j\rfloor}\varphi_{t}^{(j)}(\eta)\,\mathcal{A}_{t}(|\psi_{1}\rangle)=\sum_{t=1}^{\lfloor j\rfloor}\varphi_{t}^{(j)}(\eta)\,\mathcal{A}_{t}(|\psi_{2}\rangle).
\end{equation}

\subsection{Rotosensors for arbitrary rotation angles $\eta$ and $j\protect\leq2$
\label{subsec:jupto2}}

\subsubsection{$j=1/2$}

For a spin $1/2$, all pure states are coherent: each state $\ket\psi$ can be obtained by a suitable rotation of the state $|\tfrac{1}{2},\tfrac{1}{2}\rangle$. Since the fidelity is invariant under rotation, all states are equally sensitive to detect rotations for any angle $\eta$. 

\subsubsection{$j=1$}

For $j=1$, the expansion \eqref{PexpansionAC} takes the form 
\begin{equation}
{\cal F}_{|\psi\rangle}(\eta)=\varphi_{0}^{(j)}(\eta)+\varphi_{1}^{(1)}(\eta)\,\mathcal{A}_{1}\,,
\end{equation}
with 
\begin{equation}
\begin{aligned}\varphi_{0}^{(1)}(\eta)={} & \frac{1}{15}\big(6\cos(\eta)+\cos(2\eta)+8\big),\\
\varphi_{1}^{(1)}(\eta)={} & -\frac{1}{15}\big(2\cos(\eta)-3\cos(2\eta)+1\big).
\end{aligned}
\end{equation}
The first strictly positive zero of $\varphi_{1}^{(1)}(\eta)$ is
given by $\eta_{0}=\arccos(-2/3)$. In the interval $\eta\in[0,\eta_{0}[$,
where $\varphi_{1}^{(1)}(\eta)$ is negative, the fidelity ${\cal F}_{|\psi\rangle}(\eta)$
is minimized by states with $\mathcal{A}_{1}=1$, i.e. by $1$-anticoherent
states. For $\eta=\eta_{0}$, the fidelity takes the same value for
all states $|\psi\rangle$, namely ${\cal F}_{|\psi\rangle}(\eta_{0})=\varphi_{0}^{(1)}(\eta_{0})=7/27$.
For rotation angles in the the remaining interval, $\eta\in]\eta_{0},\pi]$,
where $\varphi_{1}^{(1)}(\eta)$ is positive, ${\cal F}_{|\psi\rangle}(\eta)$
is minimized for states with $\mathcal{A}_{1}=0$, i.e.\ coherent
states. Thus, we indeed recover the results obtained in \cite{ChrHer17}.

\subsubsection{$j=3/2$}

In this case, the average fidelity \eqref{PexpansionAC} reads 
\begin{equation}
{\cal F}_{|\psi\rangle}(\eta)=\varphi_{0}^{(3/2)}(\eta)+\varphi_{1}^{(3/2)}(\eta)\,\mathcal{A}_{1}\,,
\end{equation}
with 
\begin{equation}
\begin{aligned}\varphi_{0}^{(3/2)}(\eta)={} & \frac{1}{70}\big(\cos(3\eta)+8\cos(2\eta)+29\cos(\eta)+32\big),\\
\varphi_{1}^{(3/2)}(\eta)={} & \frac{3}{70}\big(3\cos(3\eta)+3\cos(2\eta)-4\cos(\eta)-2\big).
\end{aligned}
\end{equation}
The situation is basically the same as for $j=1$. The first strictly
positive zero of the coefficient $\varphi_{1}^{(3/2)}(\eta)$ is found
to be $\eta_{0}=\arccos(\frac{-9+\sqrt{21}}{12})$. Hence, in the
interval $\eta\in[0,\eta_{0}[$ where $\varphi_{1}^{(3/2)}(\eta)$
is negative, the fidelity ${\cal F}_{|\psi\rangle}(\eta)$ is minimal
for $1$-anticoherent states. At the value $\eta=\eta_{0}$,
the fidelity takes the same value for all states $|\psi\rangle$,
namely, ${\cal F}_{|\psi\rangle}(\eta_{0})=\varphi_{0}^{(3/2)}(\eta_{0})=(33+2\sqrt{21})/80$.
Otherwise, ${\cal F}_{|\psi\rangle}(\eta)$ is minimized for coherent
states, thereby reproducing earlier results \cite{ChrHer17}.

\subsubsection{$j=2$}

For $j=2$, the fidelity \eqref{PexpansionAC} is a linear combination
of three terms, 
\begin{equation}
{\cal F}_{|\psi\rangle}(\eta)=\varphi_{0}^{(2)}(\eta)+\varphi_{1}^{(2)}(\eta)\,\mathcal{A}_{1}+\varphi_{2}^{(2)}(\eta)\,\mathcal{A}_{2}\,,\label{Petaj2}
\end{equation}
with the angular functions $\varphi_{k}^{(2)},k=0,1,2$, displayed
in Appendix~\ref{Appendix_phi}. They all take negative values in
the interval $\eta\in[0,\eta_{0}]$, with $\eta_{0}\approx1.2122$
the first strictly positive zero of $\varphi_{1}^{(2)}(\eta)$. The tetrahedron state 
\begin{equation}
\ket{\psi^{\mathrm{tet}}}=\frac{1}{2}\left(\ket{2,-2}+i\sqrt{2}\,\ket{2,0}+\ket{2,2}\right), \label{s2}
\end{equation}
whose Majorana points lie at the vertices of a regular tetrahedron,
is $2$-anticoherent, and for $j=2$ it is the only state (up to LU)
with $\mathcal{A}_{1}=\mathcal{A}_{2}=1$~\cite{Bag14}; hence it
provides the optimal rotosensor for angles in the interval $\eta\in[0,\eta_{0}]$. For larger angles of rotation comprised between $1.68374$ and $2.44264$,
we find numerically that an optimal state is the Schrödinger cat state
\begin{equation}
\ket{\psi^{\mathrm{cat}}}=\frac{1}{\sqrt{2}}\left(\ket{2,-2}+\ket{2,2}\right)\,,\label{s2GHZ}
\end{equation}
which is only 1-anticoherent, with $\mathcal{A}_{1}=1$ and $\mathcal{A}_{2}=3/4$.
For values $\eta\gtrsim 2.44264$, the optimal state is a coherent state.

\begin{figure}[!h]
\begin{centering}
\includegraphics[width=0.475\textwidth]{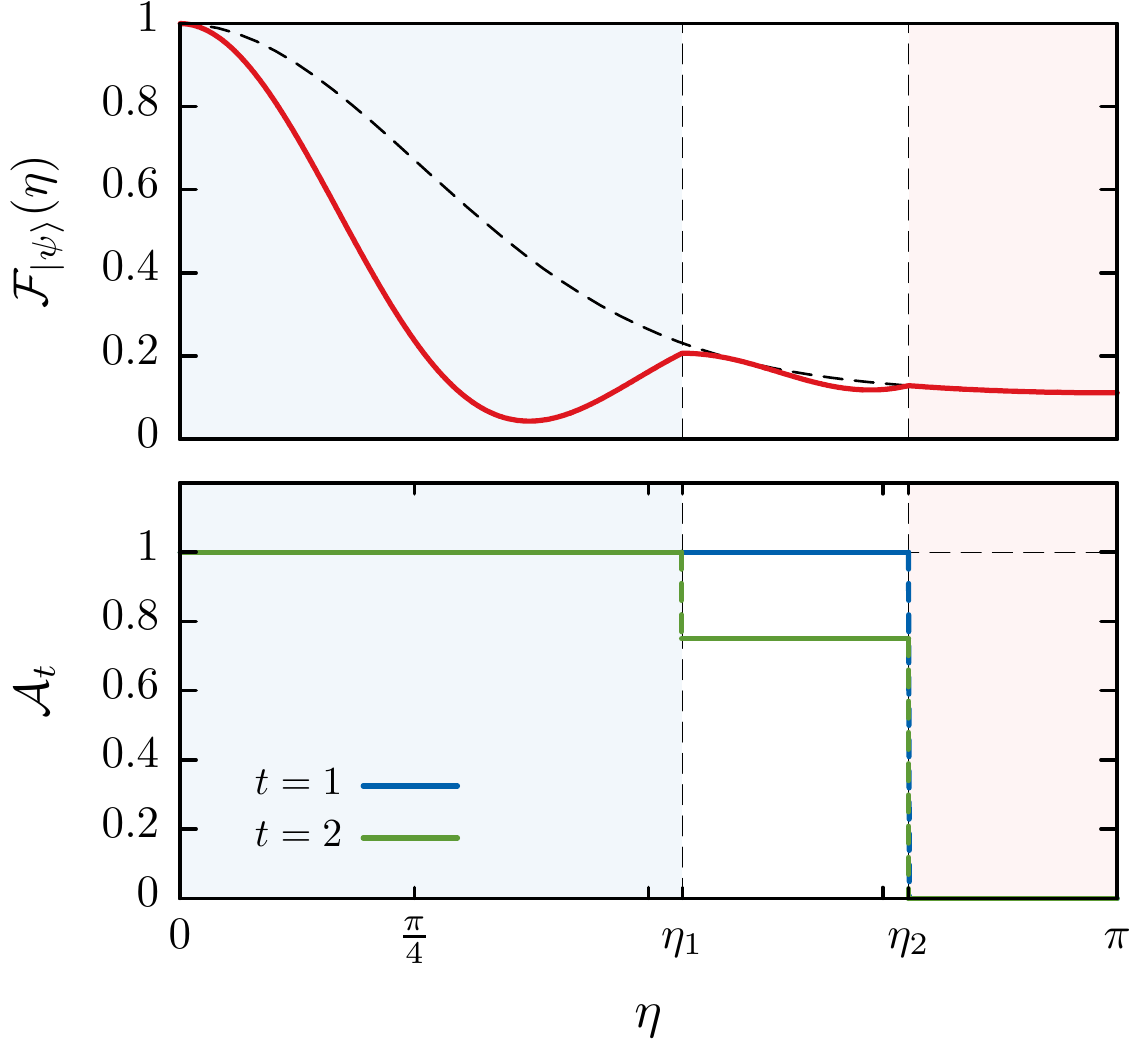} 
\par\end{centering}
\caption{Average fidelity ${\cal F}_{|\psi\rangle}(\eta)$ (top, red solid curve) and measures of anticoherence ${\cal A}_{t}$ (bottom)
for optimal states with $j=2$, as functions
of the rotation angle $\eta$; the values of the measures $\mathcal{A}_{t}$ for the optimal states are discontinuous at the values $\eta_{1}\approx 1.68374$ and $\eta_{2}\approx 2.44264$
(see text for details). The dashed curve on top shows the average fidelity $\varphi_{0}^{(2)}(\eta)$ for coherent states. The blue (red) shaded area shows the range of rotation angles for which anticoherent states to order $\lfloor j\rfloor$ (coherent states) are optimal. \label{figj2}}
\end{figure}

We thus obtain numerically three intervals with three distinct optimal states corresponding to $(\mathcal{A}_1,\mathcal{A}_2)=(1,1), (1,3/4)$, and $(0,0)$, respectively. In order to find the critical angles, we solve Eq.~\eqref{eqcritical}. The angle $\eta_{1}$ separating the first two regions is a solution of $\varphi_{2}^{(2)}(\eta)=0$. The first positive zero of $\varphi_{2}^{(2)}(\eta)$ is $\eta_{1}=2\arctan(\sqrt{9-2\sqrt{15}})\approx 1.68374$, which coincides with the numerically obtained value. The angle $\eta_{2}$ at which the second and third region touch, is a zero of $\varphi_{1}^{(2)}(\eta)+\tfrac{3}{4}\,\varphi_{2}^{(2)}(\eta)$. Its first strictly positive zero is given by
\begin{equation}
\eta_{2}=2\arctan\left(\sqrt{-\frac{a+102b}{a-38b}}\right)\,,
\end{equation}
with $a=19\ 6^{2/3}+\sqrt[3]{6}\left(223-35\sqrt{7}\right)^{2/3}$
and $b=\sqrt[3]{223-35\sqrt{7}}$, and we have indeed $\eta_{2}\approx 2.44264$.
The results we obtained are summarized
in Fig.~\ref{figj2}; they agree with the findings of~\cite{ChrHer17}.

It is noteworthy that the state \eqref{s2GHZ} is not the only state with anticoherence
measures $\mathcal{A}_{1}=1$ and $\mathcal{A}_{2}=3/4$. For instance, any state of the form
\begin{equation}
\ket{\psi}=\frac{c_1\ket{2,-1}+c_2\ket{2,0}-c_1^*\ket{2,1}}{\sqrt{2|c_1|^2+|c_2|^2}}\,
\end{equation}
with $c_1\in\mathbb{C}$ and $c_2\in\mathbb{R}$ come with the same measures of anticoherence, as readily follows from Eq.~\eqref{puritiescm}. These states are thus also optimal in the interval $\eta\in[\eta_{1},\eta_{2}]$, thereby removing the uniqueness of optimal rotosensors observed for $j=1$ and $j=3/2$.

\subsection{Rotosensors for $5/2\protect\leq j\protect\leq7/2$ \label{subsec:ju5o2pto7o2}}

\subsubsection{$j=5/2$}

For $j=5/2$, there is no anticoherent state of order $2$ but only
of order $1$~\cite{Kol08}. Numerical optimization shows that the
optimal state for small angles of rotation is the $1$-anticoherent
state with the largest measure of $2$-anticoherence, that is given
by 
\begin{equation}
\ket{\psi}=\frac{1}{\sqrt{2}}\left(\ket{\tfrac{5}{2},-\tfrac{3}{2}}+\ket{\tfrac{5}{2},\tfrac{3}{2}}\right),\label{s52}
\end{equation}
and has $\mathcal{A}_{1}=1$ and $\mathcal{A}_{2}=99/100$. This state
is found to be optimal up to a critical angle $\eta_{1}\approx1.49697$, which is obtained from Eq.~\eqref{eqcritical} and coincides with the first strictly positive zero of $\varphi_{2}^{(5/2)}(\eta)$.
It is worth noting that the optimal state \eqref{s52} was also found
to be the most non-classical spin state for $j=5/2$, both in the sense that it maximizes the quantumness~\cite{Gir10} and that it minimizes the cumulative multipole distribution~\cite{Bjo15,BjoGra15}. The Majorana constellation of this state defines a triangular bipyramid, which is a spherical $1$-design~\cite{Del77,sloane}, thus corresponding to the arrangement of point charges on the surface of a sphere which minimize the Coulomb electrostatic potential energy (solution to Thomson's problem for 5 point charges, see~\cite{Sch13}).

\begin{figure}[!h]
\begin{centering}
\includegraphics[width=0.475\textwidth]{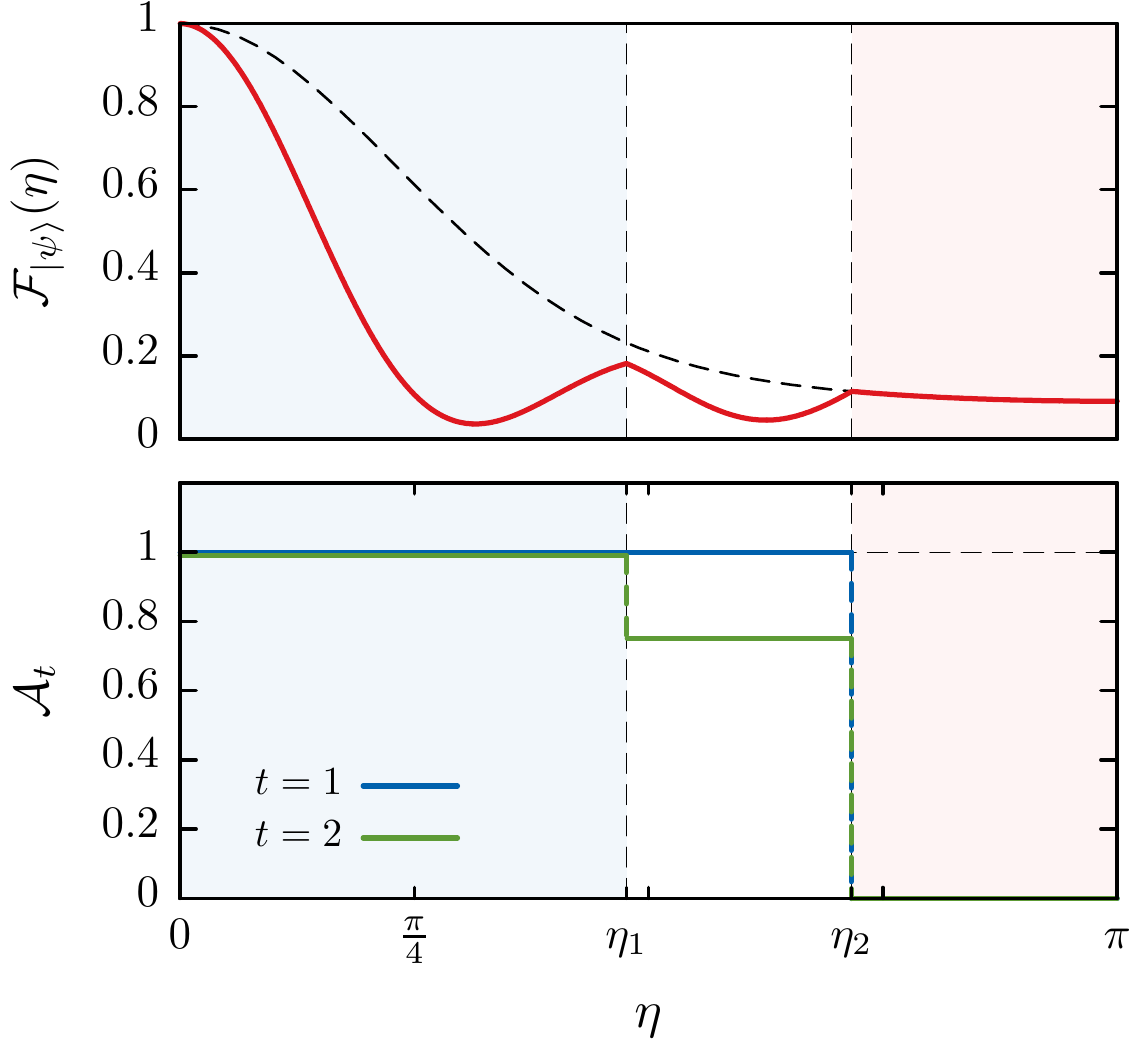} 
\par\end{centering}
\caption{Average fidelity ${\cal F}_{|\psi\rangle}(\eta)$ (top, red solid curve) and measures
of anticoherence ${\cal A}_{t}$ (bottom)
for optimal states with $j=5/2$, as functions
of the rotation angle $\eta$; the values of the measures $\mathcal{A}_{t}$ for the optimal states are discontinuous at the values $\eta_{1}\approx1.49697$ and $\eta_{2}\approx2.2521$
(see text for details). The dashed curve on top shows the average fidelity $\varphi_{0}^{(5/2)}(\eta)$ for coherent states. Shaded areas are defined as in Fig.~\ref{figj2}. \label{fig5o2}}
\end{figure}

For larger angles of rotation ranging between $\eta_{1}$ and $\eta_{2}\approx2.2521$,
we find that an optimal state is
\begin{equation}
\ket{\psi^{\mathrm{cat}}}=\frac{1}{\sqrt{2}}\left(\ket{\tfrac{5}{2},-\tfrac{5}{2}}+\ket{\tfrac{5}{2},\tfrac{5}{2}}\right)\,;\label{s5/2GHZ}
\end{equation}
unlike in the case $j=2$, we found this state for $j=5/2$ to be
the only state (up to LU) with $\mathcal{A}_{1}=1$ and $\mathcal{A}_{2}=3/4$.
For $\eta\in[\eta_{2},\pi]$, we find that coherent states are optimal.
The transition occurs at the first strictly
positive zero $\eta_{2}$ of $\varphi_{1}^{(5/2)}(\eta)+\tfrac{3}{4}\,\varphi_{2}^{(5/2)}(\eta)$.
Our results are summarized in Fig.~\ref{fig5o2}.

\subsubsection{$j=3$}

\begin{figure}[!h]
\begin{centering}
\includegraphics[width=0.475\textwidth]{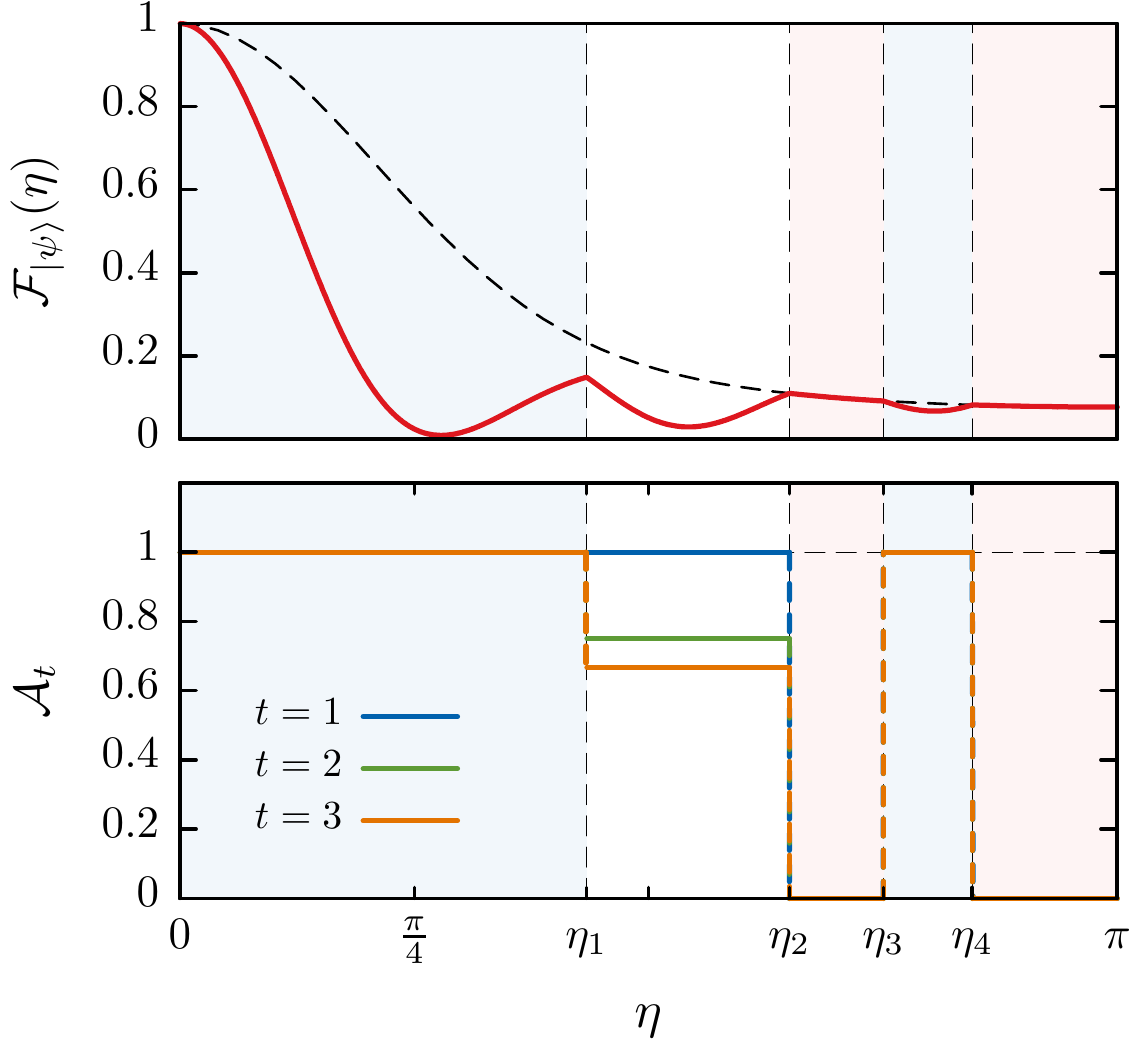} 
\par\end{centering}
\caption{Average fidelity ${\cal F}_{|\psi\rangle}(\eta)$ (top, red solid curve) and measures
of anticoherence ${\cal A}_{t}$ (bottom) 
for optimal states with $j=3$, as
functions of the rotation angle $\eta$; the values of the measures $\mathcal{A}_{t}$ for the optimal states are discontinuous at the values $\eta_{1}\approx 1.3635$, $\eta_{2}\approx 2.04367$,
$\eta_{3}\approx 2.35881$ and $\eta_{4}\approx 2.65576$ (see text
for details). The dashed curve on top shows the average fidelity $\varphi_{0}^{(3)}(\eta)$ for coherent states. Shaded areas are defined as in Fig.~\ref{figj2}. \label{figj3}}
\end{figure}

Anticoherent states of order $3$ do exist for $j=3$. They are all
connected by rotation to the octahedron state 
\begin{equation}
\ket{\psi^{\mathrm{oct}}}=\frac{1}{\sqrt{2}}\left(\ket{3,-2}+\ket{3,2}\right),\label{s3}
\end{equation}
whose Majorana points lie at the vertices of a regular octahedron.
Therefore, the state \eqref{s3} is, at small $\eta$, the unique
optimal quantum rotosensor (up to LU) for $j=3$. Numerical optimization
shows that the octahedron state is optimal up to a critical angle $\eta_{1}\approx1.3635$
coinciding with the first strictly positive zero of $\tfrac{1}{4}\,\varphi_{2}^{(3)}(\eta)+\tfrac{1}{3}\,\varphi_{3}^{(3)}(\eta)$,
and that, for larger angles, the state 
\begin{equation}
\ket{\psi^{\mathrm{cat}}}=\frac{1}{\sqrt{2}}\left(\ket{3,-3}+\ket{3,3}\right)\label{s3GHZ}
\end{equation}
with $\mathcal{A}_{1}=1$, $\mathcal{A}_{2}=3/4$ and $\mathcal{A}_{3}=2/3$
is optimal up to a critical angle $\eta_{2}\approx2.04367$ coinciding with
the first strictly positive zero of $\varphi_{1}^{(3)}(\eta)+\tfrac{3}{4}\,\varphi_{2}^{(3)}(\eta)+\tfrac{2}{3}\,\varphi_{3}^{(3)}(\eta)$.
We found that this is the only spin-$3$ state (up to LU) with $\mathcal{A}_{1}=1$,
$\mathcal{A}_{2}=3/4$ and $\mathcal{A}_{3}=2/3$. Coherent states
are found to be optimal for angles of rotation in the ranges $[\eta_{2},\eta_{3}]$
and $[\eta_{4},\pi]$ with $\eta_{3}\approx2.35881$ and $\eta_{4}\approx 2.65576$
coinciding with the second and third strictly positive zeros of $\varphi_{1}^{(3)}(\eta)+\varphi_{2}^{(3)}(\eta)+\varphi_{3}^{(3)}(\eta)$.
In the range $[\eta_{3},\eta_{4}]$, the octahedron state \eqref{s3}
becomes again optimal (although the three functions $\varphi_k^{(3)}$ for $k=1,2,3$ are not simultaneously negative in that range). Our results are displayed in Fig.~\ref{figj3}.

\subsubsection{$j=7/2$}

\begin{figure}[!h]
\begin{centering}
\includegraphics[width=0.475\textwidth]{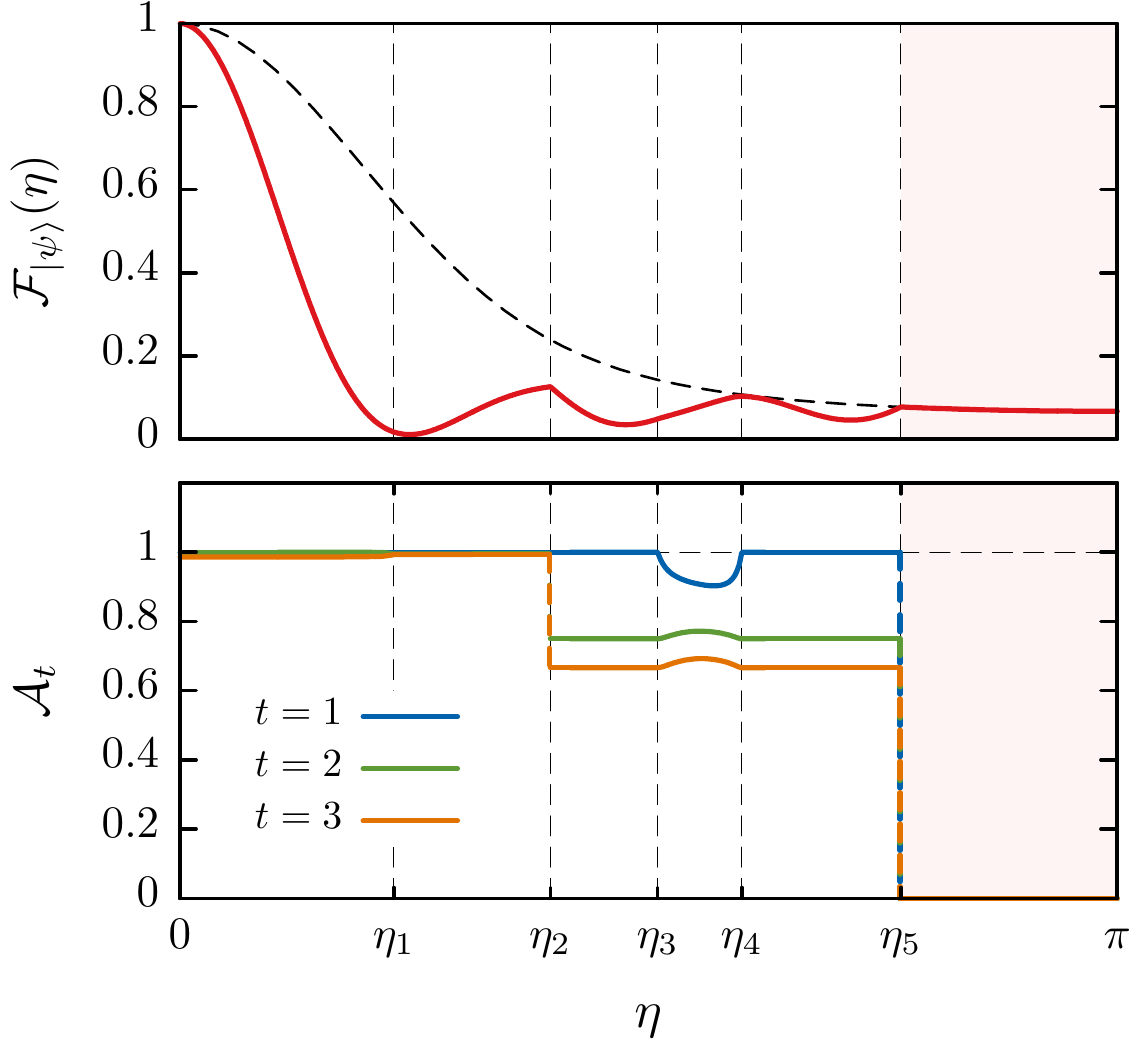} 
\par\end{centering}
\caption{Average fidelity ${\cal F}_{|\psi\rangle}(\eta)$ (top, red solid curve) and measures
of anticoherence ${\cal A}_{t}$ (bottom)
for optimal states with $j=7/2$, as functions of the rotation angle $\eta$. The dashed curve on top shows the average fidelity $\varphi_{0}^{(7/2)}(\eta)$ for coherent states. Shaded areas are defined as in Fig.~\ref{figj2}. \label{figj72}}
\end{figure}

\begin{figure}[!h]
\begin{centering}
\includegraphics[width=0.475\textwidth]{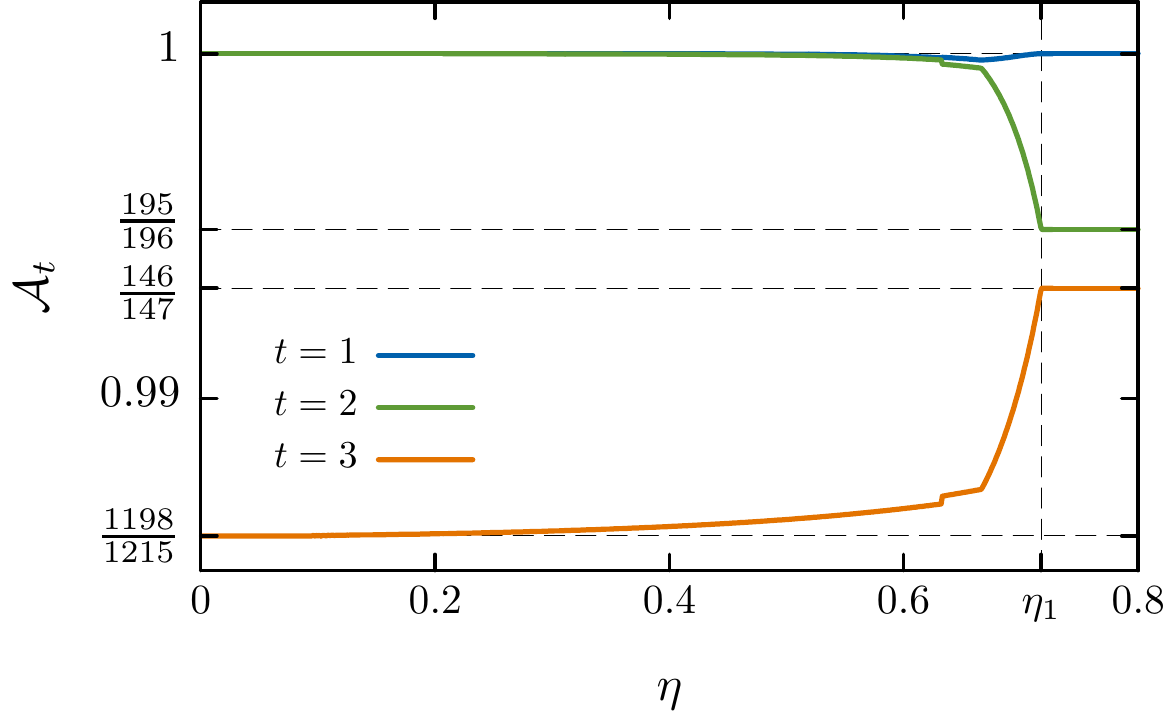} 
\par\end{centering}
\caption{Measures of anticoherence ${\cal A}_{t}$ for optimal states with $j=7/2$,
as functions of the rotation angle $\eta\in[0,0.8]$. \label{figj72bis}}
\end{figure}

This is the smallest spin quantum number for which a smooth variation
of the optimal state with $\eta$ is observed, resulting in the complex
behaviour displayed in Figs.~\ref{figj72} and \ref{figj72bis}.
There are no anticoherent states to order $3$ for $j=7/2$, but there
exist anticoherent states to order $2$. The optimal state for small
angles of rotation (by which we mean here $\eta\to0$) turns out to
be one of those. Numerical optimization yields the state 
\begin{equation}
\ket{\psi}=\sqrt{\tfrac{2}{9}}\,\ket{\tfrac{7}{2},-\tfrac{7}{2}}-\sqrt{\tfrac{7}{18}}\,\ket{\tfrac{7}{2},-\tfrac{1}{2}}-\sqrt{\tfrac{7}{18}}\,\ket{\tfrac{7}{2},\tfrac{5}{2}}\label{s72AC}
\end{equation}
with measures of anticoherence $\mathcal{A}_{1}=\mathcal{A}_{2}=1$
and $\mathcal{A}_{3}=1198/1215$. This is not the state with the highest
measure of $3$-anticoherence, as the state 
\begin{equation}
\ket{\psi}=\frac{1}{\sqrt{2}}\left(\ket{\tfrac{7}{2},-\tfrac{5}{2}}+\ket{\tfrac{7}{2},\tfrac{5}{2}}\right),
\end{equation}
has measures of anticoherence $\mathcal{A}_{1}=1$, $\mathcal{A}_{2}=195/196$
and $\mathcal{A}_{3}=146/147>1198/1215$. The latter state is found
to be optimal for $\eta\in[\eta_{1},\eta_{2}]$ with $\eta_{1}\approx0.71718$
(not identified) and $\eta_{2}\approx1.24169$ coinciding with the
first strictly positive zero of $\tfrac{12}{49}\,\varphi_{2}^{(7/2)}(\eta)+\tfrac{16}{49}\,\varphi_{3}^{(7/2)}(\eta)$.
The state 
\begin{equation}
\ket{\psi^{\mathrm{cat}}}=\frac{1}{\sqrt{2}}\left(\ket{\tfrac{7}{2},-\tfrac{7}{2}}+\ket{\tfrac{7}{2},\tfrac{7}{2}}\right)\label{s72GHZ}
\end{equation}
with $\mathcal{A}_{1}=1$, $\mathcal{A}_{2}=3/4$ and $\mathcal{A}_{3}=2/3$
is found to be optimal for $\eta\in[\eta_{2},\eta_{3}]$ and $\eta\in[\eta_{4},\eta_{5}]$
with $\eta_{3}\approx1.60141$ and $\eta_{4}\approx1.88334$ coinciding
with the third and fourth strictly positive zeros of $\varphi_{1}^{(7/2)}(\eta)$
and $\eta_{5}\approx2.41684$ with the first strictly positive zero
of $\varphi_{1}^{(7/2)}(\eta)+\tfrac{3}{4}\,\varphi_{2}^{(7/2)}(\eta)+\tfrac{2}{3}\,\varphi_{3}^{(7/2)}(\eta)$.
In the interval $[\eta_{5},\pi]$, coherent states are found to be
optimal.

\subsection{Rotosensors for small rotation angles $\eta$ and arbitrary values
of $j$ \label{subsec: Rotosensors small eta any j}}

\subsubsection{Angular functions at small angles}

According to Secs.~\ref{subsec:jupto2}
and \ref{subsec:ju5o2pto7o2} optimal
rotosensors for integer values of spin ($j=1,2,3$) are given by $j$-anticoherent
states while for half-integer spin ($j=3/2,5/2,7/2$) the fidelity
is optimized by states which are anticoherent of order $t=1,1,2$,
respectively, and possess large anticoherence measures ${\cal A}_{t}$
for values of $t$ up to $t=\left\lfloor j\right\rfloor $. This fact
can be understood quite generally through the behaviour of the functions $\varphi_{t}^{(j)}(\eta)$ at small $\eta$ for arbitrary values of $j$. In the vicinity of $\eta=0$, the functions $\varphi_{t}^{(j)}(\eta)$
given in Eq.~\eqref{Phimain} take the form 
\begin{equation}
\varphi_{t}^{(j)}(\eta)=\frac{b_{t,t}^{(j)}}{2^{2t}}\,\eta^{2t}+\mathcal{O}(\eta^{2t+2}),\label{phiseries}
\end{equation}
with coefficients $b_{t,t}^{(j)}$ given by Eq.~\eqref{btk}. These
coefficients are strictly negative for all $t\geqslant 1$ and all $j=N/2$, since $a_{t,t}^{(j)}>0$
and $a_{N-t,t}^{(j)}$ is either $0$ for $t<N/2$ or positive for
$t=N/2$. This implies that all functions $\varphi_{t}^{(j)}(\eta)$
are negative in some interval around $\eta=0$. Thus, the fidelity ${\cal F}_{|\psi\rangle}(\eta)$
is a linear combination of the $\mathcal{A}_{t}$ with negative coefficients
in that interval. Since $0\leq\mathcal{A}_{t}\leq1$, it follows that
 if there exists a state with $\mathcal{A}_{t}=1$ for all $t\leq\lfloor j\rfloor$---that
is, an anticoherent state to order $\lfloor j\rfloor$---then this
state provides an optimal quantum rotosensor for $\eta$ in that interval. 

\begin{figure}[!h]
\begin{centering}
\includegraphics[width=0.475\textwidth]{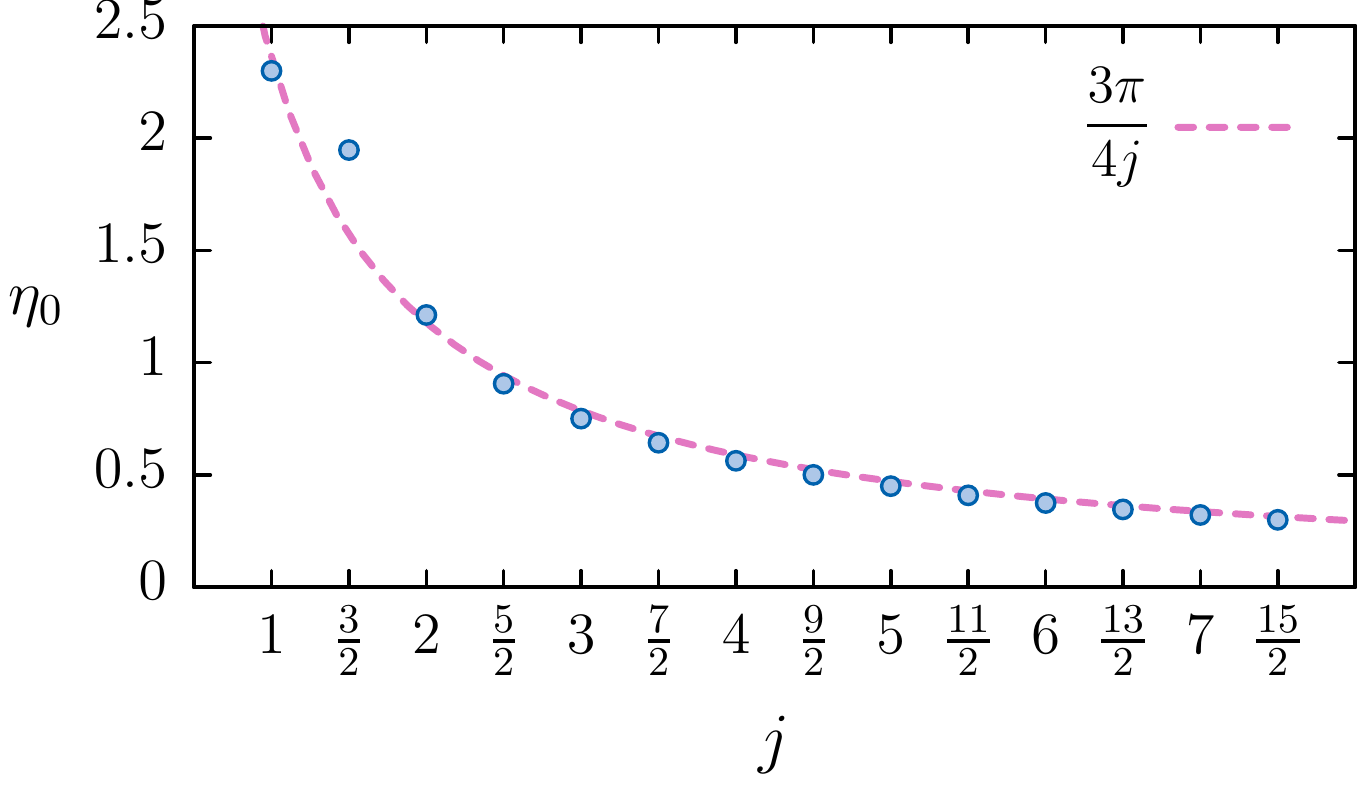} 
\par\end{centering}
\caption{First zero $\eta_{0}$ of the functions $\varphi_{1}^{(j)}(\eta)$
(blue dots) as a function of $j$: for $j=1$ and for $j\protect\geq5/2$,
the values are well approximated by $\eta_{0}\approx3\pi/(4j)$
(pink dashes).\label{etamin}}
\end{figure}

This interval can be made more specific, at least for the lowest values of $j$. Let $\eta_{0}$ denote the first zero of $\varphi_{1}^{(j)}(\eta)$. Numerical results up to $j=85$ indicate
that all functions $\varphi_{t}^{(j)}(\eta)$ for $t=1,\ldots,\lfloor j\rfloor$
are negative for $\eta\in[0,\eta_{0}]$, so that an anticoherent state to order $\lfloor j\rfloor$ (if it exists) is optimal in the whole interval $[0,\eta_{0}]$. As shown in Fig.~\ref{etamin},
$\eta_{0}$ is found to scale as $3\pi/(4j)$ for large $j$. A simple explanation for this is that the expansion of the function $\varphi_{1}^{(j)}(\eta)$ as $\sum_{k}a_{k}\cos(k\eta)$ is dominated by the term $a_{2j}\cos(2j\eta)$ (note however that $\eta_0$ is even better approximated by $9/(4j)$).
Conversely, the states maximizing ${\cal F}_{|\psi\rangle}(\eta)$
for small angles of rotation are the states with $\mathcal{A}_{t}=0$
for all $t$, i.e.~coherent states. 

To see whether any general pattern emerges, we now identify optimal
small-angle rotosensors for the next few values of the spin quantum
numbers.

\subsubsection{$j=4$}

For $j=4$, there is no anticoherent state to order $t=4$. We find
that the optimal state for small angles of rotation is the $3$-anticoherent
state 
\begin{equation}
\ket{\psi}=\sqrt{\tfrac{5}{24}}\,\ket{4,-4}-\sqrt{\tfrac{7}{12}}\,\ket{4,0}-\sqrt{\tfrac{5}{24}}\,\ket{4,4},
\end{equation}
with $\mathcal{A}_{1}=\mathcal{A}_{2}=\mathcal{A}_{3}=1$ and $\mathcal{A}_{4}=281/288$.

\subsubsection{$j=9/2$}

For $j=9/2$, there is no anticoherent state to order $t\geqslant3$.
The anticoherent states of order $t=2$ with the largest $\mathcal{A}_{3}$
are found to be of the form 
\begin{equation}
\ket{\psi}=\tfrac{\sqrt{13}}{8}\,\ket{\tfrac{9}{2},-\tfrac{9}{2}}+e^{i\chi}\sqrt{\tfrac{15}{32}}\,\ket{\tfrac{9}{2},-\tfrac{1}{2}}-\tfrac{\sqrt{21}}{8}\,\ket{\tfrac{9}{2},\tfrac{7}{2}},
\end{equation}
with $\chi\in[0,\pi/2]$. Their measures of anticoherence are $\mathcal{A}_{1}=\mathcal{A}_{2}=1$,
$\mathcal{A}_{3}=2347/2352$ and $\mathcal{A}_{4}=5\left(355609+175\sqrt{273}\cos(2\chi)\right)/1806336$.
Among these states, the one with $\chi=0$ has the largest value of
$\mathcal{A}_{4}$ and numerical results suggest that this is the
optimal state for small angles of rotation.

\subsubsection{$j=5$}

For $j=5$, there is no anticoherent state to order $t\geqslant4$.
We find that the optimal state for small angles is the $3$-anticoherent
state 
\begin{equation}
\ket{\psi}=\sqrt{\tfrac{5}{16}}\,\ket{5,-4}+\sqrt{\tfrac{3}{8}}\,\ket{5,0}-\sqrt{\tfrac{5}{16}}\,\ket{5,4},
\end{equation}
with $\mathcal{A}_{1}=\mathcal{A}_{2}=\mathcal{A}_{3}=1$, $\mathcal{A}_{4}=895/896$
and $\mathcal{A}_{5}=1097/1120$.

\begin{table*}
\begin{centering}
\begin{tabular}{|c|c|c|c|}
\hline 
$j$ & $\ket{\psi^{\mathrm{optimal}}}$ & $\mathcal{A}_{t}$ & Interval \tabularnewline
\hline 
\hline 
$1$ & $\begin{array}{c}
\ket{\psi^{\mathrm{cat}}}\\
\mathrm{any~state}\\
|j,j\rangle 
\end{array}$ & $\begin{array}{c}
\mathcal{A}_{1}=1\\
0\leqslant\mathcal{A}_{1}\leqslant 1\\
\mathcal{A}_{1}=0 
\end{array}$ & $\begin{array}{c}
\eta\in [0,\eta_{0}[ \\
\eta=\eta_{0} \\
\eta\in [\eta_{0},\pi] 
\end{array}$ \tabularnewline
\hline 
$3/2$ & $\begin{array}{c}
\ket{\psi^{\mathrm{cat}}}\\
\mathrm{any~state}\\
|j,j\rangle 
\end{array}$ & $\begin{array}{c}
\mathcal{A}_{1}=1\\
0\leqslant\mathcal{A}_{1}\leqslant 1\\
\mathcal{A}_{1}=0 
\end{array}$ & $\begin{array}{c}
\eta\in [0,\eta_{0}[ \\
\eta=\eta_{0} \\
\eta\in [\eta_{0},\pi] 
\end{array}$ \tabularnewline
\hline 
$2$ & $\begin{array}{c}
\ket{\psi^{\mathrm{tet}}}\\
\ket{\psi^{\mathrm{cat}}}\\
|j,j\rangle
\end{array}$ & $\begin{array}{c}
 \mathcal{A}_{1}=\mathcal{A}_{2}=1 \\
\mathcal{A}_{1}=1,\mathcal{A}_{2}=3/4\\
 \mathcal{A}_{1}=\mathcal{A}_{2}=0
\end{array}$ & $\begin{array}{c}
\eta\in [0,\eta_{1}], \eta_{1}\approx 1.68374\\
\eta\in [\eta_{1},\eta_{2}]\\
\eta\in [\eta_{2},\pi], \eta_{2}\approx 2.44264
\end{array}$\tabularnewline
\hline 
$5/2$ & $\begin{array}{c}
\mathrm{Eq}.~\eqref{s52}\\
\ket{\psi^{\mathrm{cat}}}\\
|j,j\rangle
\end{array}$ & $\begin{array}{c}
\mathcal{A}_{1}=1, \mathcal{A}_{2}=99/100 \\
\mathcal{A}_{1}=1,\mathcal{A}_{2}=3/4\\
\mathcal{A}_{1}=\mathcal{A}_{2}=0
\end{array}$ & $\begin{array}{c}
\eta\in [0,\eta_{1}], \eta_{1}\approx 1.49697\\
\eta\in [\eta_{1},\eta_{2}]\\\
\eta\in [\eta_{2},\pi], \eta_{2}\approx 2.2521
\end{array}$\tabularnewline
\hline 
$3$ & $\begin{array}{c}
\ket{\psi^{\mathrm{oct}}} \\
\ket{\psi^{\mathrm{cat}}}\\
|j,j\rangle
\end{array}$ & $\begin{array}{c}
\mathcal{A}_{1}=\mathcal{A}_{2}=\mathcal{A}_{3}=1 \\
\mathcal{A}_{1}=1,\mathcal{A}_{2}=3/4 ,\mathcal{A}_{3}=2/3\\
\mathcal{A}_{1}=\mathcal{A}_{2}=\mathcal{A}_{3}=0
\end{array}$ & $\begin{array}{c}
\eta\in [0,\eta_{1}]\cup [\eta_{3},\eta_{4}], \eta_{3}\approx 2.35881\\
\eta\in [\eta_{1},\eta_{2}], \eta_{1}\approx 1.3635, \eta_{2}\approx 2.04367\\
\eta\in [\eta_{2},\eta_{3}]\cup [\eta_{4},\pi], \eta_{4}\approx 2.65576
\end{array}$\tabularnewline
\hline 
$7/2$ & $\begin{array}{c}
\mathrm{Eq}.~\eqref{s72AC} \\
-  \\
\ket{\psi^{\mathrm{cat}}}\\
- \\
|j,j\rangle
\end{array}$ & $\begin{array}{c}
\mathcal{A}_{1}=\mathcal{A}_{2}=1, \mathcal{A}_{3}=1198/1215 \\
\tfrac{195}{196}\leqslant\mathcal{A}_{2}\leqslant 1,  \tfrac{1198}{1215}\leqslant\mathcal{A}_{3}\leqslant \tfrac{146}{147},\, \mathrm{see~Fig.}~\ref{figj72bis} \\
\mathcal{A}_{1}=1,\mathcal{A}_{2}=3/4 ,\mathcal{A}_{3}=2/3\\
\mathrm{see~Fig.}~\ref{figj72} \\
\mathcal{A}_{1}=\mathcal{A}_{2}=\mathcal{A}_{3}=0
\end{array}$ & $\begin{array}{c}
\eta \to 0 \\
\eta\in [0,\eta_{1}],\eta_{1}\approx 0.71718\\
\eta\in [\eta_{2},\eta_{3}]\cup [\eta_{4},\eta_{5}], \eta_{2}\approx 1.24169\\
\eta\in [\eta_{3},\eta_{4}], \eta_{3}\approx 1.60141, \eta_{4}\approx 1.88334\\
\eta\in [\eta_{5},\pi], \eta_{5}\approx 2.41684
\end{array}$\tabularnewline
\hline 
\end{tabular}
\caption{Summary of the results of Secs.~\ref{subsec:jupto2} and \ref{subsec:ju5o2pto7o2} on optimal states for $1\leq j \leq 7/2$. Here, $\eta_{0}$ denotes the first strictly positive zero of $\varphi_{1}^{(j)}(\eta)$, $\ket{\psi^{\mathrm{tet}}}$ defined for $j=2$ is given by Eq.~\eqref{s2}, $\ket{\psi^{\mathrm{oct}}}$ defined for $j=3$ is given by Eq.~\eqref{s3}, and $\ket{\psi^{\mathrm{cat}}}=\frac{1}{\sqrt{2}}\left(\ket{j,-j}+\ket{j,j}\right)$ for any $j$. The state $|j,j\rangle$ has been taken as an example of coherent state. Note that optimal states given here are not necessarily unique (states not related by a rotation can have the same $\mathcal{A}_t$). \label{tab}}
\par\end{centering}
\end{table*}

\subsubsection{Arbitrary values of $j$}

As was mentioned earlier, if an anticoherent state to order $\lfloor j\rfloor$
exists for a given $j$, then this state gives rise to an optimal quantum
rotosensor for $\eta\in[0,\eta_{0}]$. This applies to values $j=1,3/2,2$
and $j=3$, which are the only cases where existence of anticoherent
states to order $t=\lfloor j\rfloor$ has been established (see e.g.~\cite{Bag15,Bag17}).

The situation is less straightforward if such a state is not known to exist from the outset. The only general conclusion one can draw is that minimizing the average fidelity ${\cal F}_{|\psi\rangle}(\eta)$  for a fixed angle $\eta\in[0,\eta_{0}]$ corresponds to maximizing the measures $\mathcal{A}_{t}$ within the domain $\Omega$ (by definition, $\Omega$ is the set of all reachable $\mathcal{A}_{t}$ so that by changing $|\psi\rangle$, we will remain within $\Omega$). In this sense, the more anticoherent a state is, the more sensitive it will be as a quantum rotosensor. In general, varying $|\psi\rangle$ will change all anticoherence measures simultaneously.  The challenge is to determine whether a state with given values of the measures $\mathcal{A}_{t}$ exists and, if it does, to identify it.

The maximal order of anticoherence that a spin-$j$ state can display
is generally much smaller than $\lfloor j\rfloor$, typically $t\sim2\sqrt{j}$
for large spins $j$~\cite{Bag15}. Numerical results for $j\lesssim100$
seem to suggest that the pairs $(t,j)$ for which a $t$-anticoherent
spin-$j$ state exists coincide with those for which a $2j$-points
spherical $t$-design exists in three dimensions~\cite{Grashttp}.
The latter have been tabulated up to $j=50$ \cite{sloane}. For example,
the first pairs $(t,j)$ for $j\leq4$ are given by $(1,1)$, $(1,3/2)$,
$(2,2)$, $(1,5/2)$, $(3,3)$, $(2,7/2)$, $(3,4)$.

\section{Summary and conclusions \label{sec:Conclusion}}

The main result of this work is a closed-form expression
\eqref{PexpansionAC} for the fidelity ${\cal F}_{|\psi\rangle}(\eta)$
between a state and its image under a rotation by an angle $\eta$
about an axis ${\bf n}$, averaged over all rotation axes. The expression
takes the form of a linear combination of anticoherence measures $\mathcal{A}_{t}$, with explicit $\eta$-dependent coefficients. It follows that not
only spin-$j$ states which are related by a global rotation of the axes come with the same average fidelity, but more generally all states with identical purities of their reduced density matrices (calculated for any subset of their $2j$ constituent spin-$1/2$ in the Majorana representation). This gives an explanation for the observation of~\cite{ChrHer17} that optimal states are not necessarily unique. Moreover, since the fidelity is linear in the anticoherence measures, optimal states correspond to values of $\mathcal{A}_{t}$ on the boundary of the domain $\Omega$ of admissible values. This shows the relevance of characterizing the domain $\Omega$.

The expression \eqref{PexpansionAC} allows us to
characterize states which optimally detect rotations by their degree
of coherence or anticoherence. At small angles $\eta\leq\eta_{0}$,
where the coefficients of the measures $\mathcal{A}_{t}$ are all
negative, optimality of detection of rotations goes hand in hand with
high degrees of anticoherence. For angles close to $\eta=\pi$, however,
numerical results support the claim that optimality is achieved throughout
by spin coherent states. 

We also performed a systematic investigation of
states minimizing the average fidelity for small values of $j$, for
all integers and half-integers from $j=1/2$ to $j=5$. Table~\ref{tab} summarizes our findings for the lowest values of $j$. At small rotation
angle, all optimal states were found to have a maximal lowest
anticoherence measure: $\mathcal{A}_{1}=1$. These states, which are
anticoherent to order $1$, exist for any value of $j$, and one may
conjecture that they should, in fact, be optimal for arbitrary values
of $j$. More generally, for all values of $j$ investigated and for
$\eta\leq\eta_{0}$, the optimal states turned out to have, for each
$t>1$, the largest admissible anticoherence measure $\mathcal{A}_{t}$
compatible with fixed values of the lower measures $\mathcal{A}_{1},\mathcal{A}_{2},\ldots,\mathcal{A}_{t-1}$.
Whether this property holds in general remains an open question.

Note that natural generalizations of this problem, such as maximization of the average fidelity, can also be addressed by our approach. For instance, for small rotation angles $\eta\in [0,\eta_0]$, where all $\varphi_t^{(j)}(\eta)$ with $t\geqslant 1$ are negative, the average fidelity is maximal for coherent states. For rotation angles close to $\eta=\pi$, numerical results indicate that the $1$-anticoherent state $\ket{\psi^{\mathrm{cat}}}=\frac{1}{\sqrt{2}}\left(\ket{j,-j}+\ket{j,j}\right)$ is optimal for all $j$ up to $17/2$.

\begin{acknowledgments}
OG and SW thank the hospitality of the University of Liège, where
this work has been initiated.
\end{acknowledgments}

\appendix

\section{Fidelity in parameter estimation theory of rotations}
\label{Appendix_param}
It was shown in~\cite{Goldberg18} that minimizing the uncertainty in the measurement
of an \emph{unknown} angle about a \emph{known} rotation axis is equivalent
to identifying the states which minimize the fidelity $F_{|\psi\rangle}(\eta,\mathbf{n})$,
assuming small rotation angles and using parameter estimation theory.
To see this, first expand the fidelity as~\begin{equation}
|\bra\psi R_{\mathbf{n}}(\eta)|\psi\rangle|^{2}=1-\eta^{2}(\Delta J_{\mathbf{n}})^{2}+{\cal O}(\eta^{4})\,,\label{eq: Goldberg fidelity}
\end{equation}
where $(\Delta J_{\mathbf{n}})^{2}$ is the variance of $J_{\mathbf{n}}\equiv\mathbf{J}\boldsymbol{\cdot}\mathbf{n}$
in the state $|\psi\rangle$. Solving \eqref{eq: Goldberg fidelity}
for the angle $\eta$ will, upon measuring the fidelity in any state
$|\psi\rangle$, result in an estimate of $\eta$. Second, the accuracy
of this value depends on the initial state $|\psi\rangle$: using
error propagation, one finds that the variance of the estimator is approximately
given by 
\begin{equation}
\left(\Delta\eta\right){}^{2}\approx\frac{1}{\left(2\Delta J_{{\bf n}}\right)^{2}}\,.\label{eq: angle variance}
\end{equation}
Thus, states $|\psi\rangle$ with \emph{large }values of the variance
\emph{$\left(\Delta J_{{\bf n}}\right)^{2}$} are seen to minimize
the uncertainty of the angle $\eta$. According to Eq. \eqref{eq: Goldberg fidelity}, these states also \emph{minimize} the fidelity $F_{|\psi\rangle}(\eta,\mathbf{n})$.

Let us generalize the argument to the case in which the rotation axis
is \emph{unknown}. We will see that the states producing the most
reliable results---i.e.\ the smallest variance in the angle estimator---are
those which minimize the average fidelity ${\cal F}_{|\psi\rangle}(\eta)$.
It is convenient to describe the randomness in the rotation axis in
terms of a quantum channel (see for instance \cite{Sidhu+2019} for the use of channels in quantum estimation theory). Suppose we prepare the pure
initial state $\rho_{0}=\ket\psi\bra\psi$ and send it through the
$\eta$-dependent channel $\Lambda_{\eta}(\cdot)$,
\begin{equation}
\rho_{\eta}=\Lambda_{\eta}(\rho_{0})=\frac{1}{4\pi}\int_{{{\cal S}^2}}R_{\mathbf{n}}(\eta)\rho_{0}R_{\mathbf{n}}^{\dagger}(\eta)\,d{\bf n}\,,\label{eq: rho eta}
\end{equation}
which describes rotations by $\eta$ about all possible rotation axes.
Next, we measure the projector $\rho_{0}=\ket\psi\bra\psi$.
Assuming the rotation angle to be small, $\eta\ll1$, the
probability to still find the propagated state $\rho_{\eta}$ in the
initial state $\rho_{0}$ is given by
\begin{equation}
\langle\rho_{0}\rangle_{\eta}=\tr\left[\rho_{0}\rho_{\eta}\right]=1-\eta^{2}V+{\cal O}(\eta^{4})\,,\label{eq: mean of rho_0}
\end{equation}
where $V$ is the variance of $J_{\mathbf{n}}\equiv\mathbf{J}\boldsymbol{\cdot}\mathbf{n}$,
averaged over all directions, 
\begin{equation}
V=\frac{1}{4\pi}\int_{{\cal S}^2}\left(\bra\psi J_{\mathbf{n}}^{2}\ket\psi-\bra\psi J_{\mathbf{n}}\ket\psi^{2}\right)\,d{\bf n}\,.\label{eq: average variance}
\end{equation}
 Using $\rho_{0}^{2}=\rho_{0}$ and the relation \eqref{eq: mean of rho_0},
the variance of the measurement outcomes is found to
be 
\begin{equation}
\left(\Delta\rho_{0}\right)_{\eta}^{2}=\tr\left[\rho_{0}^{2}\rho_{\eta}\right]-\left(\tr\left[\rho_{0}\rho_{\eta}\right]\right)^{2}=\eta^{2}V+{\cal O}(\eta^{4})\,.\label{eq: variance of rho_0}
\end{equation}
Now using again the error propagation formula instrumental in the
derivation of Eq.~\eqref{eq: angle variance} about \emph{known }rotation
axes\emph{, }its generalization to \emph{unknown }axes is given by
\begin{equation}
\left(\Delta\eta\right){}^{2}=\frac{\left(\Delta\rho_{0}\right)_{\eta}^{2}}{|\partial\langle\rho_{0}\rangle_{\eta}/\partial\eta|^{2}}+{\cal O}(\eta^{2})\approx\frac{1}{4V}\,.\label{eq: eta variance (unknown axes)}
\end{equation}
 This result concludes the argument we wish to provide: it is of
physical interest to minimize the average fidelity 
\begin{equation}
{\cal F}_{|\psi\rangle}(\eta)\equiv\tr\left[\rho_{0}\rho_{\eta}\right]\approx1-\eta^{2}V\,,\label{eq: average fidelity (alternative)}
\end{equation}
since the states which do so are those states which allow one to
most accurately estimate a (small) rotation angle about unknown axes.

\section{Average fidelity for Dicke states\label{sec: appendix C (Dicke)}}

For Dicke states $|j,m\rangle$ (common eigenstates of $\mathbf{J}^{2}$ and
$J_{z}$), the average fidelity \eqref{eq: probability} reads
\begin{equation}
\begin{aligned}{\cal F}_{|j,m\rangle}(\eta) & {}=\frac{1}{4\pi}\int_{\unitS}|\bra{j,m}R_{\mathbf{n}}(\eta)\ket{j,m}|^{2}\,d\mathbf{n}\\
 & {}=\frac{1}{4\pi}\int_{\unitS}|U_{mm}^{j}(\eta,\mathbf{n})|^{2}\,d\mathbf{n}
\end{aligned}
\end{equation}
with $U_{mm}^{j}(\eta,\mathbf{n})\equiv U_{mm}^{j}$ a matrix element of the rotation operator in the
angle-axis parametrization given by
\begin{equation}
U_{mm}^{j}=\frac{\sqrt{4\pi}}{2j+1}\sum_{\lambda,\mu}(-i)^{\lambda}\sqrt{2\lambda+1}\chi_{\lambda}^{j}(\eta)C_{jm\lambda\mu}^{jm}Y_{\lambda}^{m}(\mathbf{n})\label{Umm}
\end{equation}
where $C_{jm\lambda\mu}^{jm}$ are Clebsch-Gordan coefficients, $Y_{\lambda}^{m}(\mathbf{n})$
are spherical harmonics and $\chi_{\lambda}^{j}(\eta)$ are the generalized
characters of order $\lambda$ of the irreducible representations
of rank $j$ of the rotation group~\cite{Var88}. These are defined
by 
\begin{equation}
\chi_{\lambda}^{j}(\eta)=\sqrt{\tfrac{(2j+1)(2j-\lambda)!}{(2j+\lambda+1)!}}\sin^{\lambda}\left(\tfrac{\eta}{2}\right)\left(\tfrac{d}{d\cos\left(\tfrac{\eta}{2}\right)}\right)^{\lambda}\chi^{j}(\eta)\label{chilj}
\end{equation}
with the characters 
\begin{equation}
\chi^{j}(\eta)=\frac{(4j+2)!!}{2(4j+1)!!}\,P_{2j}^{\big(\tfrac{1}{2},\tfrac{1}{2}\big)}\left(\cos\left(\tfrac{\eta}{2}\right)\right)
\label{chij}
\end{equation}
where $P_{n}^{(\alpha,\beta)}$ are Jacobi polynomials. Taking the
modulus squared of \eqref{Umm} and integrating over all directions
by using orthonormality of the spherical harmonics, we readily get
Eq.~\eqref{PDicke}.

\section{Explicit calculation of the angular functions $\varphi_{t}^{(j)}(\eta)$
\label{appexplicit}}

\subsection{Matrices $S_{\mu_{1}\mu_{2}\ldots\mu_{N}}$}

The matrices $S_{\mu_{1}\mu_{2}\ldots\mu_{N}}$ with $N=2j$ appearing in the
expansion \eqref{rhoarbitrary} can be obtained by expanding the $(j,0)$
representation of a Lorentz boost, 
\begin{equation}
\Pi^{(j)}(q)\equiv(q_{0}^{2}-|\bq|^{2})^{j}\,e^{-2\theta_{q}\,\hbq\boldsymbol{\cdot}\bJ},\label{Aplorentzboost}
\end{equation}
with $\theta_{q}=\mathrm{arctanh}(-|\bq|/q_{0})$ and $\hbq=\bq/|\bq|$.
This expansion takes the form of a multivariate polynomial in the
variables $q_{0},q_{1},q_{2},q_{3}$, 
\begin{equation}
\Pi^{(j)}(q)=(-1)^{2j}q_{\mu_{1}}q_{\mu_{2}}\ldots q_{\mu_{2j}}S_{\mu_{1}\mu_{2}\ldots\mu_{2j}},
\label{Apegaliteweinberg}
\end{equation}
where the coefficients are the $(N+1)\times(N+1)$ matrices $S_{\mu_{1}\mu_{2}\ldots\mu_{N}}$~\cite{prl}.

\subsection{Tensor coordinates of the maximally mixed state}

The maximally mixed state $\rho_{0}=\idmat/(N+1)$ can be expanded
along \eqref{rhoarbitrary} with coefficients $x_{\mu_{1}\mu_{2}\ldots\mu_{N}}^{(0)}$. The coherent state decomposition of the maximally
mixed state, $\rho_{0}=\frac{1}{4\pi}\int_{\unitS}\ket{\mathbf{n}}\bra{\mathbf{n}}\,d\mathbf{n}$,
yields the identity 
\begin{equation}
x_{\mu_{1}\mu_{2}\ldots\mu_{N}}^{(0)}=\frac{1}{4\pi}\int_{\unitS}n_{\mu_{1}}n_{\mu_{2}}\ldots n_{\mu_{N}}\,d\mathbf{n}.\label{identiterho0}
\end{equation}

Using our convention not to write indices when they are equal to 0,
we have, irrespective of spin size, $x_{0}^{(0)}=1$, $x_{aa}^{(0)}=1/3$,
$x_{aaaa}^{(0)}=1/5$ and $x_{aabb}^{(0)}=1/15$ for $a\neq b$. More
generally, the coefficients of the maximally mixed state are given
by the polynomial identity (cf. Eq.~(27) of \cite{prl}) 
\begin{equation}
x_{\mu_{1}\mu_{2}\ldots\mu_{N}}^{(0)}q_{\mu_{1}}\ldots q_{\mu_{N}}=\sum_{k=0}^{j}\frac{\binom{N}{2k}}{2k+1}q_{0}^{N-2k}|\bq|^{2k}\,,\label{coorid}
\end{equation}
which leads to 
\begin{equation}
x_{a_{1}a_{2}\ldots a_{N}}^{(0)}=\frac{1}{N+1}\frac{\binom{j}{p_{1}/2,p_{2}/2,p_{3}/2}}{\binom{N}{p_{1},p_{2},p_{3}}}\,,\label{coorid2}
\end{equation}
where $p_{i}$ denotes the number of $i$ in $\{a_{1},a_{2},\ldots,a_{N}\}$ and the terms in the fraction are multinomial coefficients (by convention the right-hand side evaluates to zero if some $p_{i}$
is not even).

\subsection{Average fidelity in terms of tensor coordinates}

According to Eq.~\eqref{Ptot-1}, the average fidelity can be written
as a double sum, 
\begin{equation}
\begin{aligned} & {\cal F}_{|\psi\rangle}(\eta)=\sum_{k=0}^{N}(-1)^{N}\frac{q_{0}^{2(N-k)}}{m^{2N}}\\
 & \times\hspace{-0.5cm}\sum_{\genfrac{}{}{0pt}{1}{\boldmu\bm{,}\bn}{2(N-k)\textrm{zeros}}{}}\hspace{-0.5cm}(-1)^{\textrm{nr of 0 in }\bn}x_{\mu_{1}\ldots\mu_{N}\nu_{1}\ldots\nu_{N}}^{(0)}x_{\mu_{1}\ldots\mu_{N}}x_{\nu_{1}\ldots\nu_{N}}\,.
\end{aligned}
\label{Ptot}
\end{equation}
We now wish to show that the second sum which runs over all strings
of indices (between 0 and 3) containing $2(N-k)$ zeros can evaluated
explicitly leading to the simpler form for ${\cal F}_{|\psi\rangle}(\eta)$
given in Eq.~\eqref{Ptot6} at the end of this section.

The sum runs over terms containing $2(N-k)$ zeros, that is, $2k$
non-zero indices. We split it into terms containing $r$ nonzero indices
in $\boldmu$ and $2k-r$ in $\bn$. At fixed $k$ we have 
\begin{align}
 & \sum_{\genfrac{}{}{0pt}{1}{\boldmu\bm{,}\bn}{2(N-k)\textrm{zeros}}{}}\hspace{-0.5cm}(-1)^{\textrm{nr of 0 in }\bn}x_{\mu_{1}\ldots\mu_{N}}x_{\nu_{1}\ldots\nu_{N}}x_{\mu_{1}\ldots\mu_{N}\nu_{1}\ldots\nu_{N}}^{(0)}\nonumber \\
 & =\sum_{r=2k-N}^{N}(-1)^{N-2k+r}\binom{N}{r}\binom{N}{2k-r}\times\nonumber \\
 & \qquad\times\sum_{a_{i},b_{i}}x_{a_{1}...a_{r}}x_{b_{1}...b_{2k-r}}x_{a_{1}...a_{r}b_{1}...b_{2k-r}}^{(0)}.\label{eq: fixed k sum}
\end{align}

We now evaluate the sums $\sum_{a_{i},b_{i}}x_{a_{1}...a_{r}}x_{b_{1}...b_{2k-r}}x_{a_{1}...a_{r}b_{1}...b_{2k-r}}^{(0)}$.
We may suppose that $r\leqslant2k-r$. Using \eqref{coorid2}, we
see that the nonzero indices $a_{i}$ and $b_{i}$ must occur in pairs.
Indices $a_{i}$ are either paired with indices $a_{k}$ or indices
$b_{k}$. We can then split the sum according to the number of pairings
of the form $(a_{i},b_{i})$ (all other pairings are then within the
$a_{i}$ or within the $b_{i}$). Let us first consider the case $k=r$. We are going to show that
\begin{equation}
\begin{aligned}
\sum_{a_{i},b_{i}} &{} x_{a_{1}...a_{r}}x_{b_{1}...b_{r}}x_{a_{1}...a_{r}b_{1}...b_{r}}^{(0)} =\\
 &{} \lambda_{0}\sum_{a_{i}}x_{a_{1}...a_{r}}^{2} \\
&{} + \lambda_{1}\sum_{a_{i}}\left(\sum_{b}x_{a_{1}...a_{r-2}bb}\right)^{2} \\
 &{} + \lambda_{2}\sum_{a_{i}}\left(\sum_{b_{1},b_{2}}x_{a_{1}...a_{r-4}b_{1}b_{1}b_{2}b_{2}}\right)^{2}+\ldots
\end{aligned}\label{kequalr}
\end{equation}
with 
\begin{equation}
\lambda_{q}=\frac{2^{r-2q}r!^{2}}{(2r+1)!}\binom{r}{r-2q,q,q}.
\end{equation}
We first use the explicit expression \eqref{coorid2} of $x_{a_{1}...a_{r}b_{1}...b_{r}}^{(0)}$ to get an equation equivalent to \eqref{kequalr}, namely
\begin{align}
 & \sum_{c_{i}}x_{c_{1}...c_{r}}x_{c_{r+1}...c_{2r}}\frac{\binom{2r}{r}\binom{r}{p_{1}/2\,p_{2}/2\,p_{3}/2}}{\binom{2r}{p_{1}\,p_{2}\,p_{3}}}=2^{r}\sum_{a_{i}}x_{a_{1}...a_{r}}^{2}\nonumber \\
 & +2^{r-2}\binom{r}{r-2,1,1}\sum_{a_{i}}\left(\sum_{b}x_{a_{1}...a_{r-2}bb}\right)^{2}+\cdots\nonumber \\
 & +2^{r-2q}\binom{r}{r-2q,q,q}\sum_{a_{i}}\left(\sum_{b}x_{a_{1}...a_{r-2q}b_{1}b_{1}...b_{q}b_{q}}\right)^{2}\nonumber \\
 & +\cdots,\label{kequalr2}
\end{align}
where $p_{i}$ is the number of $i$ in $\{c_{1},c_{2},\ldots,c_{2r}\}$
and terms with $p_{i}$ odd are zero. In order to prove Eq.~\eqref{kequalr2}, we just observe 
that it represents two different ways of counting the same quantity. Indeed,
let $\eta_{i}=\{a_{i},\epsilon_{i},\epsilon_{i}'\}$ for $1\leqslant i\leqslant r$
be triplets with $1\leqslant a_{i}\leqslant3$ and $0\leqslant\epsilon_{i},\epsilon_{i}'\leqslant1$.
To a given set $\mathbf{\eta}=\{\eta_{1},\ldots,\eta_{r}\}$ we associate a term
of the form $x_{c_{1}...c_{r}}y_{c_{r+1}...c_{2r}}$ where the $c_{i}$
occur in pairs $(a_{1},a_{1}),(a_{2},a_{2}),\ldots,(a_{r},a_{r})$.
In a pair $(a_{i},a_{i})$, the first $a_{i}$ is assigned to be an index
of $x$ if $\epsilon_{i}=0$, of $y$ if $\epsilon_{i}=1$ (and similarly
the second $a_{i}$ in the pair is an index of $x$ if $\epsilon_{i}'=0$,
of $y$ otherwise). For instance, $\eta=(a,0,0)$ corresponds to a term $x_{aa\ldots}y_{\ldots}$ and
$\eta=(a,0,1)$ corresponds to a term $x_{a\ldots}y_{a\ldots}$. In order that $x$ and $y$ have the same number $r$ of indices we need to have $\sum_{i}(\epsilon_{i}+\epsilon_{i}')=r$, so that among the 
$\epsilon_{i},\epsilon_{i}'$ there are $r$ 0's and $r$ 1's.
Each $\mathbf{\eta}=\{\eta_{1},\ldots,\eta_{r}\}$ such that $\sum_{i}(\epsilon_{i}+\epsilon_{i}')=r$ then corresponds to a unique term of the form $x_{c_{1}...c_{r}}y_{c_{r+1}...c_{2r}}$.
Consider now, for some $q\leq r$, all $\mathbf{\eta}$ with $\sum_{i}(\epsilon_{i}+\epsilon_{i}')=r$ for which $\epsilon_{i}=\epsilon_{i}'=0$ for exactly $q$ values of $i$.
These correspond to terms $x_{c_{1}...c_{r}}y_{c_{r+1}...c_{2r}}$ such that exactly $q$ pairs $(a_i,a_i)$ appear as indices of $x$, $q$ pairs appear as indices of $y$, and $r-2q$ are distributed over $x$ and $y$, i.e.~terms of the form $x_{a_{1}a_2...a_{r-2q}b_1b_1 b_2b_2\ldots b_qb_q}y_{a_{1}a_2...a_{r-2q}c_1c_1 c_2c_2\ldots c_qc_q}$.
Replacing $y$ by $x$, all these terms are those appearing in the right-hand side of \eqref{kequalr2}. In fact, each sum on the right-hand side of \eqref{kequalr2} can be interpreted as the sum over
all $\eta_{i}$ such that $\sum_{i}(\epsilon_{i}+\epsilon_{i}')=r$
and $\epsilon_{i}=\epsilon_{i}'=0$ for exactly $q$ values of $i$. 
For instance the first term on the
right-hand side of \eqref{kequalr2} corresponds to terms $q=0$, where all
pairs $(a_{i},a_{i})$ are distributed over the two different strings
of indices (and then of course replacing $y$ by $x$). The prefactors correspond to the ways of choosing the positions of a given set of pairs: the multinomial coefficient corresponds to the choice of positions of the indices among the $r$ indices of $x_{a_{1}a_2...a_{r-2q}b_1b_1 b_2b_2\ldots b_qb_q}$. The factor $2^{r-2q}$ corresponds to choosing between $x$ and $y$ for the $r-2q$ indices $a_i$ which are distributed over $x$ and $y$.

The same sum can be expressed as the left-hand-side of
\eqref{kequalr2} if we now first sum over all strings $c_{1}\leqslant c_{2}\leqslant\cdots\leqslant c_{2r}$,
which implies dividing by the number of permutations $\binom{2r}{p_{1},p_{2},p_{3}}$,
then consider all possible positions of the $a_{i}$ over the $r$
pairs, which implies multiplying by the number of permutations of
the pairs $\binom{r}{p_{1}/2,p_{2}/2,p_{3}/2}$, and finally choose
the $r$ entries among the $\epsilon_{i}$ and $\epsilon_{i}'$ that
will take the value 0, hence the factor $\binom{2r}{r}$. Thus \eqref{kequalr2}
holds, which proves \eqref{kequalr}. The tracelessness condition \eqref{traceless} then allows us to reduce
the sums over $b$ in \eqref{kequalr} to invariants $\kappa_{r}$. We simply get
\begin{equation}
\begin{aligned} & \sum_{a_{i},b_{i}}x_{a_{1}...a_{r}}x_{b_{1}...b_{r}}x_{a_{1}...a_{r}b_{1}...b_{r}}^{(0)}\\
 & \qquad=\lambda_{0}\kappa_{r}+\lambda_{1}\kappa_{r-2}+\lambda_{2}\kappa_{r-4}+\ldots
\end{aligned}
\label{eq: reduce sum}
\end{equation}

Following exactly the same procedure from \eqref{kequalr} to \eqref{eq: reduce sum} in the case where the strings of indices of $x$ and $y$ have different lengths, we obtain the more general expression
\begin{align}
 & \sum_{a_{i},b_{i}}x_{a_{1}...a_{r}}x_{b_{1}...b_{2k-r}}x_{a_{1}...a_{r}b_{1}...b_{2k-r}}^{(0)}\nonumber \\
 & =\frac{r!(2k-r)!}{(2k+1)!}\sum_{q=0}^{\lfloor\frac{r}{2}\rfloor}2^{r-2q}\binom{k}{r-2q,q,q+k-r}\kappa_{r-2q}.\label{kequalrgen}
\end{align}
From \eqref{Ptot} we finally get 
\begin{widetext}
\begin{equation}
\begin{aligned}
{\cal F}_{|\psi\rangle}(\eta)={}&\sum_{k=0}^{N}(-1)^{k}\sin^{2k}\left(\frac{\eta}{2}\right)\cos^{2(N-k)}\left(\frac{\eta}{2}\right)\sum_{r=0}^{2k}(-1)^{r}\frac{N!^{2}}{(N-r)!(N-2k+r)!(2k+1)!}\\
& \times \sum_{q=0}^{\lfloor\frac{r}{2}\rfloor}2^{r-2q}\binom{k}{r-2q,q,q+k-r}\kappa_{r-2q}.
\end{aligned}\label{Ptot2}
\end{equation}
Changing the summation over $r$ to a summation over $s=r-2q$, we get
\begin{equation}
\begin{aligned}
{\cal F}_{|\psi\rangle}(\eta)={}&\sum_{k=0}^{N}(-1)^{k}\sin^{2k}\left(\frac{\eta}{2}\right)\cos^{2(N-k)}\left(\frac{\eta}{2}\right)
\sum_{s=0}^{2k}(-2)^{s}\kappa_{s} \\
& \times\sum_{q=0}^{k-\lfloor\frac{s+1}{2}\rfloor} \binom{N}{s+2q} \binom{N}{2k-2q-s} \frac{(s+2q)!(2k-2q-s)!}{(2k+1)!}\binom{k}{s,q,k-q-s}.
\end{aligned}\label{Ptot2b}
\end{equation}
Because of the multinomial coefficient at the end of  \eqref{Ptot2b}, the sum over $s$ can be restricted to $s\leqslant k$ and the sum over $q$ to $q\leqslant k-s$, yielding
\begin{equation}
\begin{aligned}
{\cal F}_{|\psi\rangle}(\eta)={}&\sum_{k=0}^{N}(-1)^{k}\sin^{2k}\left(\frac{\eta}{2}\right)\cos^{2(N-k)}\left(\frac{\eta}{2}\right)\frac{N!^{2}k!}{(2k+1)!}\\
& \times \sum_{s=0}^{k}\frac{(-2)^{s}}{s!(2N-2k)!(k-s)!}\sum_{q=0}^{k-s}\binom{2N-2k}{N-s-2q}\binom{k-s}{q}\kappa_{s}.
\end{aligned}\label{Ptot3}
\end{equation}
Grouping the $\kappa_{s}$ together by changing the order of the sum
we get 
\begin{equation}
\begin{aligned}
{\cal F}_{|\psi\rangle}(\eta)={}& N!^{2}\sum_{s=0}^{N}\frac{(-2)^{s}\kappa_{s}}{s!}\sum_{k=s}^{N}(-1)^{k}\sin^{2k}\left(\frac{\eta}{2}\right)\cos^{2(N-k)}\left(\frac{\eta}{2}\right)\frac{k!}{(2k+1)!}\\
& \times \sum_{q=0}^{k-s}\frac{1}{(N-s-2q)!(N-2k+s+2q)!(k-s-q)!q!}.
\end{aligned}\label{Ptot4}
\end{equation}
Because of the sum over $q$ from 0 to $k-s$, we can make the sum
over $k$ start at 0. We then use \eqref{invrel} to express the $\kappa_{s}$
in terms of $\tr\left[\rho_{t}^{2}\right]$. This gives
\begin{equation}
\begin{aligned}
{\cal F}_{|\psi\rangle}(\eta) ={}& N!^{2}\sum_{t=0}^{N}\frac{(-2)^{t}}{t!}\tr\left[\rho_{t}^{2}\right]\sum_{k=0}^{N}(-1)^{k}\sin^{2k}\left(\frac{\eta}{2}\right)\cos^{2(N-k)}\left(\frac{\eta}{2}\right)\frac{k!}{(2k+1)!}\\
 & \times\sum_{s=t}^{N}\frac{2^{s}}{(s-t)!}\sum_{q=0}^{k-s}\frac{1}{(N-s-2q)!(N-2k+s+2q)!(k-s-q)!q!}. 
\end{aligned}\label{Ptot5}
\end{equation}
It turns out that the sums in the second line of this expression can
be performed. Indeed, the identity 
\begin{equation}
\sum_{s=t}^{N}\frac{2^{s}}{(s-t)!}\sum_{q=0}^{k-s}\frac{1}{(N-s-2q)!(N-2k+s+2q)!(k-s-q)!q!}=\frac{2^{t}(2N-2t)!}{(N-t)!^{2}(k-t)!(2N-2k)!}\label{idcompliquee}
\end{equation}
holds for arbitrary $N,t,k$. This can be proved as follows. First
change variables $N\to N-t$, $k\to k-t$ and $s\to s-t$, so that
showing \eqref{idcompliquee} amounts to showing 
\begin{equation}
\sum_{s=0}^{k}\frac{2^{s}}{s!}\sum_{q=0}^{k-s}\frac{1}{(N-s-2q)!(N-2k+s+2q)!(k-s-q)!q!}=\frac{(2N)!}{N!^{2}k!(2N-2k)!}\label{idcompliquee2}
\end{equation}
(the upper bound of the sum over $s$ can be changed from $N$ to
$k$ since terms $s>k$ do not contribute). Equation~\eqref{idcompliquee2}
can be rewritten 
\begin{equation}
\sum_{s=0}^{k}\sum_{q=0}^{k-s}2^{s}\binom{k}{s}\binom{k-s}{q}\binom{2N-2k}{N-s-2q}=\binom{2N}{N}.\label{idcompliquee3}
\end{equation}
Such an identity can be proven by writing $(1+x)^{2N}=(1+2x+x^{2})^{k}(1+x)^{2N-2k}$
for any $k$ and any $x$, and expanding the first factor using multinomial
coefficients and the second one using binomial coefficients: 
\begin{align*}
(1+x)^{2N}= & (1+2x+x^{2})^{k}(1+x)^{2N-2k}\\
 & =\sum_{s,q}\binom{k}{s,q,k-s-q}(2x)^{s}(x^{2})^{q}\sum_{u}\binom{2N-2k}{u}x^{u}\\
 & =\sum_{s,q,u}2^{s}\binom{k}{s}\binom{k-s}{q}\binom{2N-2k}{u}x^{u+s+2q}
\end{align*}
(the boundaries of the sums are taken care of by the binomial coefficients
which vanish outside a certain range of parameters). Identifying the
coefficients of the term in $x^{N}$ readily gives \eqref{idcompliquee3}.

Using \eqref{idcompliquee}, Eq.~\eqref{Ptot5} finally reduces to
\begin{equation}
{\cal F}_{|\psi\rangle}(\eta)=\frac{1}{2N+1}\frac{1}{\binom{2N}{N}}\sum_{t=0}^{N}(-4)^{t}\binom{2N-2t}{N-t}\tr\left[\rho_{t}^{2}\right]\sum_{k=0}^{N}(-1)^{k}\sin^{2k}\left(\frac{\eta}{2}\right)\cos^{2(N-k)}\left(\frac{\eta}{2}\right)\binom{2N+1}{2k+1}\binom{k}{t}.\label{Ptot6}
\end{equation}

\section{Angular functions for $j=2$}
\label{Appendix_phi}

Evaluating the expression \eqref{Phimain} for $j=2$ leads to these
three angular functions:
\begin{equation}
\begin{aligned}
\varphi_{0}^{(2)}(\eta)={} & \frac{1}{315}\left(130\cos(\eta)+46\cos(2\eta)+10\cos(3\eta)+\cos(4\eta)+128\right),\\
\varphi_{1}^{(2)}(\eta)={} & -\frac{4}{315}\left(10\cos(\eta)-11\cos(2\eta)+16\cos(3\eta)-20\cos(4\eta)+5\right),\\
\varphi_{2}^{(2)}(\eta)={} & -\frac{64}{105}\sin^{4}\left(\frac{\eta}{2}\right)(10\cos(\eta)+5\cos(2\eta)+6)\,.
\end{aligned}
\label{varphij2}
\end{equation}

\section{Sample code}
\label{code}

We give here a short sample code written in Mathematica\textsuperscript{\texttrademark} to find an optimal state for $j=5/2$ and $\eta=0.5$.\vspace*{2pt}

\begin{lstlisting}[language=Mathematica, mathescape]
(* Angular functions, see Eqs. (42), (45) and (44) *)
a[n_,t_,k_]:=((-1)^(k+t)*4^t)/(2*k+1)*(Binomial[2*n,2*k]*Binomial[2*n-2*t,n-t] Binomial[k,t])/Binomial[2*n,n];
b[n_,t_,k_]:=If[t==0,Binomial[n,k]/(1+2*k),-(t/(t+1))*If[t==n/2,1/2,1]*(a[n,t,k]+a[n,n-t,k])];
phi[n_,t_,eta_]:=Sum[Sin[eta/2]^(2*k)*Cos[eta/2]^(2*(n-k))*b[n,t,k],{k,0,n}];

(* Measures of anticoherence of a pure state of the form (25), 
   see Eqs. (10) and (27) *)
A[cm__,n_,t_]:=(t+1)/t*(1-Sum[Abs[Sum[Conjugate[cm[[k+l+1]]]*cm[[k+q+1]]*Sqrt[Binomial[k+l,k]*Binomial[n-k-l,t-l]*Binomial[k+q,k]*Binomial[n-k-q,t-q]]/Binomial[n,t],{k,0,n-t}]^2],{q,0,t},{l,0,t}])

(* Average fidelity, see Eq. (3) *)
F[cm__,n_,eta_]:=phi[n,0,eta]+Sum[phi[n,t,eta]*A[cm,n,t],{t,1,Floor[n/2]}];

(* Normalized state expressed in the Dicke basis for j=5/2 *)
j=5/2; n=2*j;
cm=Normalize@(Array[r,n+1,0]+I*Array[i,n+1,0]);

(* Search for an optimal state for eta=0.5 *)
eta=0.5;
f=Simplify[ComplexExpand[F[cm,n,eta]]];
sol=NMinimize[f,Array[r,n+1,0]~Join~Array[i,n+1,0],AccuracyGoal->25,PrecisionGoal->25];

(* Minimal average fidelity *)
Re@sol[[1]]

(* Optimal state in the Dicke basis *)
cmsol=cm/.sol[[2]]

(* Measures of anticoherence of order 1 and 2 *)
A[cmsol,n,1]
A[cmsol,n,2]

\end{lstlisting}

\end{widetext}

\noindent The evaluation of the code with Mathematica 12.0 yields the output
\begin{lstlisting}[language=Mathematica, mathescape]
0.453337

{0.189461+0.48194 I, -0.155904-0.0488534 I, 0.0828666+0.00440845 I, 0.374917+0.583967 I, 0.257018-0.0504367 I, -0.165557+0.347373 I}

1.

0.99
\end{lstlisting}

The state that is found, with measures of anticoherence $\mathcal{A}_{1}=1$ and $\mathcal{A}_{2}=99/100$, can be shown to be related by a rotation to the state \eqref{s52}.

\end{document}